\documentclass[traditabstract]{aa} 
\usepackage{graphicx}
\usepackage{longtable}
\usepackage{amsfonts}
\usepackage{rotating}
\RequirePackage{amssymb}
\RequirePackage{latexsym}

\usepackage{natbib}

\def\bmy{\hbox{\it b--y\/~}}
\def\feh{\hbox{\rm [Fe/H]~}}
\def\mh{\hbox{\rm [M/H]~}}
\def\zx{\hbox{\rm Z/X}}

\def\afe{\hbox{\rm [$\alpha$/Fe]~}}
\renewcommand{\min}{\mbox{$^m$}}

\def\min{${}^{\prime}$}

\newcommand{\strom}{\mbox{Str\"omgren~}}
\newcommand{\stromc}{\mbox{Str\"omgren}}

\usepackage{txfonts}
\begin{document}
   \title{\strom-- near-infrared photometry of the Baade's Window. I. 
The bulge globular cluster NGC~6528 and the surrounding field.}

\titlerunning{\strom-- near-infrared photometry of the Baade's Window}
\authorrunning{Calamida et al.}

   \author{A. Calamida\inst{2, 3}
   \and G. Bono\inst{3, 4}
   \and E. P. Lagioia\inst{5}
   \and A. P. Milone\inst{6}
   \and M. Fabrizio\inst{7}
   \and I. Saviane\inst{8}
   \and C. Moni Bidin\inst{9}
   \and F. Mauro\inst{10}
   \and R. Buonanno\inst{4}
   \and I. Ferraro\inst{3}
   \and G. Iannicola\inst{3}
   \and M. Zoccali\inst{11}}

   \institute{
   Based on observations collected with EFOSC2@NTT (Program ID:085.D-0374).\\
   \and Space Telescope Science Institute, 3700 San Martin Drive, Baltimore, MD 21218 \\
   \email{calamida@stsci.edu}
   \and INAF-Osservatorio Astronomico di Roma, Via Frascati 33, 00040, Monte Porzio Catone, Italy \\
   \email{annalisa.calamida@oa-roma.inaf.it; ivan.ferraro@oa-roma.inaf.it; giacinto.iannicola@oa-roma.inaf.it}
   \and Universit\`a di Roma Tor Vergata, Via della Ricerca Scientifica 1, 00133 Rome, Italy \\
   \email{Giuseppe.Bono@roma2.infn.it; roberto.buonanno@roma2.infn.it}
   \and Dipartimento di Fisica e Astronomia, Universit\`a di Bologna, Viale Berti Pichat, 6/2, 40127, Bologna, Italy\\
   \email{edoardo.lagioia2@unibo.it}
   \and Research School of Astronomy and Astrophysics, The Australian National University, Cotter Road, Weston, ACT, 2611, Australia\\
   \email{milone@mso.anu.edu.au}
   \and INAF-Osservatorio Astronomico di Collurania, Teramo\\ 
   \email{fabrizio@oa-teramo.inaf.it}
   \and  ESO Chile, Santiago, Chile\\
   \email{isaviane@eso.org}
   \and Instituto de Astronom\'ia, Universidad Cat\'olica del Norte, Av, Angamos 0610, Antofagasta, Chile\\
   \email{cmoni@ucn.cl}
   \and Universidad de Concepci\'on, Concepci\'on, Chile\\
   \email{fmauro@astroudec.cl}
   \and Pontificia Universidad Cat\'olica de Chile, Av. Vicuna Mackenna 4860, 782-0436 Macul, Santiago, Chile\\
   \email{mzoccali@astro.puc.cl}}

   \date{}

 
  \abstract
   {We present \strom and near-infrared (NIR) photometry of the bulge cluster NGC~6528 and
   its surrounding field in the Baade's Window. $uvby$ images have been collected with EFOSC2 
   on the New Technology Telescope (NTT, La Silla, ESO). 
   The NIR catalogs are based on $J,K$-band VIRCAM@VISTA (Paranal, ESO) and SOFI@NTT photometry. 
   The matching of the aforementioned data sets with Hubble Space Telescope photometry allowed us to 
   obtain proper-motion-cleaned samples of NGC~6528 and bulge stars. 
   Furthermore, we were able to correct the \stromc--NIR photometry for differential reddening.
   The huge color sensitivity of the \stromc--NIR Color--Magnitude--Diagrams (CMDs) helped us 
   in disentangling age and metallicity effects. The red-giant branch (RGB) of NGC~6528 is well reproduced 
   in all the CMDs by adopting scaled-solar isochrones with solar abundance, i.e. $Z = 0.0198$, 
   or $\alpha$-enhanced isochrones with the same iron content, i.e. $Z = 0.04$, and an age range of 
   $t$ = 10 - 12 Gyr. The same isochrones well reproduce most of the color spread of 
   the Baade's window RGB. These findings support age estimates present in literature
   for NGC~6528.
   
   We also performed a new theoretical visual--NIR metallicity calibration based 
   on the \strom index $m_1$ and on visual--NIR colors for red-giant (RG) stars. 
   Scaled-solar and $\alpha$-enhanced models have been adopted and we validated the
   new Metallicity--Index--Color (MIC) relations by applying them to estimate the photometric
   metal abundance of a sample of field RGs and of a metal-poor (M~92, \feh $\sim -2.3$) and a
   metal-rich (NGC~6624, \feh $\sim -0.7$) globular cluster.
   We applied the calibration to estimate the mean metal abundance of NGC~6528, finding
   \feh = \mh = $-0.04\pm$0.02, with a mean intrinsic dispersion of $\sigma =$ 0.27 dex, by  
   averaging the metallicities obtained with the scaled-solar $[m],\ y - J$ and $[m],\ y - K$
   MIC relations, and of $-0.11\pm$0.01, with $\sigma =$  0.27 dex, 
   by using the $m_1,\ y - J$ and $m_1,\ y - K$ relations. 
   These findings would support the results of Zoccali et al. (2004) based on high-resolution spectroscopy, 
   which give \feh = $-0.10\pm$0.2 for NGC~6528, and a low $\alpha$-enhancement of \afe = 0.1, 
   and of Carretta et al. (2001), that find  \feh = 0.07$\pm$0.01, with a modest $\alpha$-enhancement, 
   \afe = 0.2. By applying the scaled-solar MIC relations to the sample of Baade's window RGs, 
   we find a metallicity distribution which extend approximately from \feh $\sim -1.0$ up to \feh $\sim$ 1 dex,
   with two metallicity peaks at \feh $\approx -0.2$ and \feh $\approx$ 0.55 ($[m],\ y - J$  and $[m],\ y - K$ relations),
   and \feh $\approx -0.25$ and \feh $\approx$ 0.4 ($m_1,\ y - J$ and $m_1,\ y - K$ relations).
   These findings are in fairly good agreement with the spectroscopic studies of Zoccali et al.\ (2008), 
   and Hill et al. (2011) for the Baade's window, of Uttenthaler et al. (2012) for a region centered at 
   $(l,b) = (0^{\circ}, -10^{\circ})$, and with the recent results of the ARGOS surveys for the bulge 
   (Freeman et al.\ 2013, Ness et al.\ 2013a).}
   
   \keywords{stars: abundances --- stars: evolution -- Galaxy: globular clusters}

   \maketitle
%

\section{Introduction}
The study of the Galactic bulge is fundamental to understand the star formation history of the Galaxy. 
Moreover, the Milky Way bulge is the closest galaxy bulge that can be observed, and constraining 
its properties is of great help to the study of more distant galaxies. 
The Galactic bulge is, indeed, dominated by metal-rich old stars ($t > 10$ Gyr, \citealt{zocc03}), 
but the presence of a few intermediate-age stars is not excluded \citep{vanloon03, bensby13}. 
The metallicity distribution of red-giant (RG) stars in the bulge spans a range from \feh $\approx -1.5$ to 0.5, 
according to high-resolution spectroscopy of $\sim$ 800 stars \citep[hereafter ZO08]{hill11, zocc08}. 
Furthermore, \citet{oscar} and \citet{utten} recently showed that bulge stars 
are $\alpha$-enhanced when compared with thin disk stars. Moreover, there is 
evidence that the  $\alpha$ enhancement decreases at higher metallicities. 
More recently, the ARGOS (Abundances and Radial velocity Galactic Origins Survey) 
survey provided medium-resolution spectroscopy for $\sim$ 25,000 red clump stars 
distributed in 28 two-degree fields at different Galactic longitudes and for 
latitudes $b = -5, -7.5, -10^{\circ}$. 
They found that the bulge stellar content is made by two spatially and 
chemically distinct sub--populations.
{\em a) \/} A thin more metal-rich sub--sample, peaking at super-solar iron abundance 
(\feh $\approx$ 0.15), that is kinematically colder and closer to the 
Galactic plane (A component). This component is associated with the thin disc of the Galaxy.
{\em b) \/}  A thick more metal-poor sub-sample (peaking at \feh $\approx -0.25$)
that is uniformly distributed across the selected fields. This component
is associated with the boxy/peanut bulge \citep{ness13}.
They also identify a main more metal-poor component at \feh = $-0.66$, 
and two other peaks in their metallicity distribution function (MDF), 
at \feh $=  -1.16$, and $-1.73$, that appear to be associated 
with the thick disk and with the inner halo \citep{freeman13, ness13}.
These results are also in agreement with the high-resolution 
spectroscopy of a sample of 58 micro-lensed bulge dwarfs by 
\citet{bensby13}. Their MDF presents three main "bumps" at \feh $\approx -0.7, -0.25$, and
0.3, which are in agreement with the three main populations identified by the ARGOS survey.
The above MDFs agree quite well with the results by \citet{hill11} for the Baade's
window (BW). They found, by using high-resolution spectra for 219 bulge red clump giants,
a bi-modal metallicity distribution, with peaks at \feh $\sim -0.3$ and 0.3. 
The two main components seem to show different kinematical properties,
with the most metal--poor stars more compatible with an old-spheroid, while 
the metal-rich component with a bar population. 
A double peaked MDF was also found by \citet{utten}, based on medium-resolution 
spectra of $\sim$ 400 RGs in a bulge field centered at $(l,b) = (0^{\circ}, -10^{\circ})$.
The two peaks are located at \mh $\sim -0.6$ and $\sim$ 0.3, suggesting that their 
metal-poor metallicity peak is shifted towards the more metal-poor regime when  
compared to the metal-poor peaks found by \citet{bensby11} and \citet{hill11}.
Moreover, they also found that the metal-rich stars show a smaller velocity 
dispersion compared to metal-poor stars. On the basis of this evidence they suggest 
that the metal-rich population, making up $\approx$ 30\% of the sample, 
could belong to the bar, while the metal-poor one to the Galactic bulge.

A quantitative explanation of how the Galactic bulge formed is still missing, 
and medium- and high-resolution spectroscopic studies are tracing different 
regions of the bulge. Some evidence points to constrain the Milky Way bulge 
as a classical bulge, others favors a pseudo-bulge. The fact that the 
RR Lyrae do not show evidence of a bar is also suggesting that the truly 
old (horizontal branch stars) bulge population forms a spherical bulge, 
while the intermediate-age population (red clump stars) form a boxy/peanut 
bulge \citep{vasquez, dekany}.

In this context, an interesting region to constrain the bulge nature, 
is the BW. It is characterized by a low-reddening, 
$E(B-V) <$ 0.6 mag, and by the presence of two Galactic globular clusters 
(GGCs), namely NGC~6522 and NGC~6528. The former is classified as 
a bulge metal-intermediate cluster (\feh $\sim -1.3/-1.2$ dex).
\citet{io11} estimated a mean metal-abundance of \feh $= -1.0$ for this cluster,
by adopting a \strom theoretical metallicity calibration based on the $hk$ index. 
The latter --NGC~6528-- is among the most metal-rich GGCs.
Recent spectroscopic investigations, based on high-resolution spectra, 
suggested for this cluster a solar metallicity and a modest $\alpha$-enhancement. 
\citet{carretta01}, by using four red horizontal-branch (RHB) stars, 
found \feh = 0.07$\pm$0.01 and an $\alpha$-enhancement of \afe $\approx$ 0.2, 
while \citet{zocc04}, by using three RGB stars found \feh = $Ð0.1\pm$0.2 and \afe = 0.1$\pm$ 0.1. 
More recently \citet{origlia}, using high-resolution NIR spectra of four bright giants, 
found \feh =  $Ð0.17\pm$0.01 and a higher $\alpha$-enhancement, 
\afe = 0.33$\pm$0.01.
The quoted measurements indicate that NGC~6528 is an ideal laboratory 
not only to constrain the $\alpha$-enhancement in old metal-rich systems, 
but also to shed new light on the possible occurrence of an age--metallicity 
relation among the most metal-rich Galactic globulars 
(Rakos \& Schombert 2005; Dotter et al.\ 2011).

The metallicity estimates, based on photometric indices, of field and cluster 
stars in the bulge are partially hampered by the occurrence of differential 
reddening across the field of view and by the mix of stellar populations
belonging to the bulge, the thick and thin disk and to globulars. 
The first problem can be partially overcome by using NIR bands, while 
the latter is more complex. The use of optical--NIR color--color planes, 
in particular \stromc--NIR planes, to separate cluster and field stars 
\citep{io09, bono10} is limited by the fact that the metallicity distributions 
of bulge, thick/thin disk and NGC~6528 all peak around solar chemical composition. 
Furthermore, the use of the color--color plane to separate field and 
cluster stars does require precise and deep photometry in at least three 
optical--NIR bands. 

In the current paper we adopt ground--based photometry and use the 
\stromc--NIR color--color planes to perform a preliminary selection of bulge, 
disk and cluster stars. We also take advantage of optical and NIR space photometry 
collected with Advanced Camera for Surveys (ACS) and Wide Field Camera 3 (WFC3) on board 
the Hubble Space Telescope (HST) to estimate proper motions and further select the different 
stellar components. The ground--based \stromc--NIR photometry is then adopted 
to estimate the metallicity distribution of bulge and cluster RGs, 
by applying a new theoretical calibration of the \strom $m_1 = (v-b) - (b-y)$ metallicity index, based
on visual--NIR colors.

This is the first paper of a series devoted to \stromc--NIR photometry of metal--rich GGCs.
In this initial investigation, we derive and test a new theoretical calibration of the 
$m_1$ index, based on visual--NIR colors for RG stars. The new calibration is 
adopted to constrain the metallicity distribution of NGC~6528 and its surrounding 
field.

We favored the adoption of the visual--NIR colors to derive Metallicity--Index--Color (MIC) relations, 
since they have two clear advantages when compared with indices only based on \strom colors.
{\em a) \/} They are not hampered by the presence of $CN$,$CH$, and $NH$ 
molecular absorption bands. Two strong $CN$ bands ($\lambda=4142$ and $\lambda=4215$ \AA) 
affects the  $v$ filter, while the strong $CH$-band at $\lambda=4300$ \AA~ might affect 
both the $v$ and the $b$, and the $NH$-band at $\lambda=3360$ \AA~ plus
the two $CN$ bands at $\lambda=3590$ and $\lambda=3883$ \AA~ might affect the $u$;
{\em ib) \/} They have a stronger sensitivity to effective temperature. 
The quoted molecular bands still affect the the new visual--NIR MIC relations, 
but only through the \strom $m_1$ index.

The new MIC relations are based on scaled-solar and $\alpha$-enhanced evolutionary models. 
The adopted evolutionary models were transformed into the observational plane by adopting  
bolometric corrections  and color--temperature transformations based on atmosphere models 
constructed by adopting the same chemical mixtures. They are valid in the metallicity 
range -2.5 $ < \feh <$ 0.5. 
The new MIC relations were validated by estimating the abundances of GGCs, 
from the very metal-poor M~92 (\feh $\sim -2.3$) to the bulge metal-rich cluster 
NGC~6624 (\feh $\sim -0.6$), and of field RG stars, and by comparing
them with medium- and high-resolution spectroscopic measurements.

The structure of the current paper is as follows. 
In \S 2 we discuss in detail the observations and data reduction, while in \S 3 we describe 
the photometric calibration. In \S 4 we present the \stromc--NIR catalogs adopted 
in this paper.
Section 5 deals with the approach adopted to calibrate the visual--NIR MIC relations
for giant stars, while in \S 6 we present the different tests we performed to validate the current 
theoretical MIC relations together with the comparison between photometric estimates 
and spectroscopic measurements of metal abundances. 
In \S 7 we study the metallicity distribution of NGC~6528 and the Baade's window RGs. 
The summary of the results and a brief discussion concerning further developments 
of the new MIC relations are given in \S 8.

\section{Observations and data reduction}

We adopt \strom photometry of the bulge cluster NGC~6528 collected with 
the EFOSC2 camera on the NTT in 2010 (Program ID: 085.D-0374).
The CCD is a Loral/Lesser Thinned AR coated 2048$\times$2048 chip, with a pixel size of 15 $\mu$m,
corresponding to a pixel scale of 0\farcs 12. 
The field of view (FoV) is thus 4.1\min$\times$4.1\min. We used the normal read-out mode and 
the 2$\times$2 binning, getting an effective pixel scale of 0\farcs 24. 

We collected 2$u$-, 2$v$-, 3$b$- and 3$y$-band images during two different nights (July 11 and 13), 
with exposure times ranging from $60$ to $2,500$s and seeing from 0\farcs7 to 1\farcs0. 
The total FoV is $\approx$ 5\min$\times$5\min and includes the cluster center. 
The footprints of the observed fields are shown in Fig.~1 (blue), while
the log of the observations is given in Table~1.

\begin{figure}
 \includegraphics[width=0.5\textwidth]{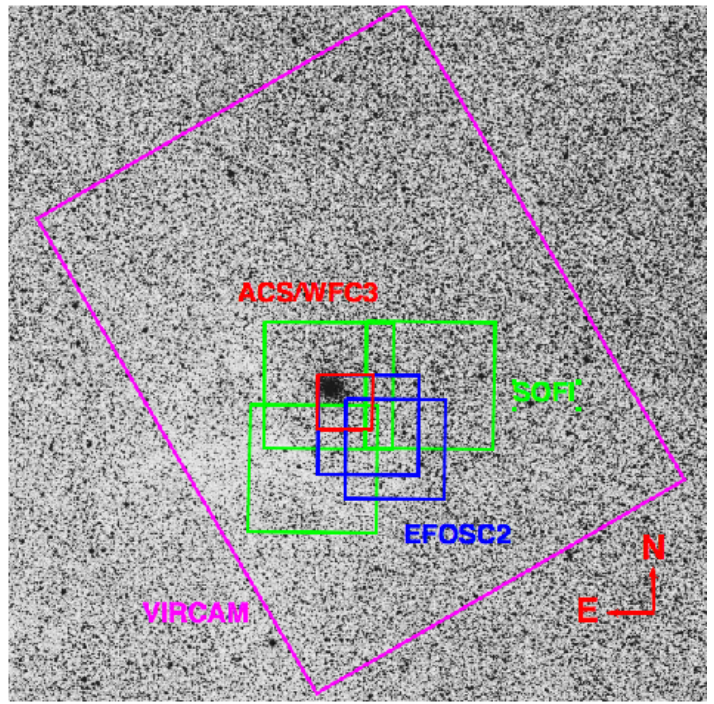}
      \caption{Footprints of the different optical -- NIR data sets adopted in the study of the
      bulge cluster NGC~6528 and its surrounding field.}
          \label{fig1}
   \end{figure}
   
Halo clusters for which we have published \strom photometry \citep[hereafter CA07]{gru00,gru02,io07} 
were observed during the same nights with similar airmass conditions. 
The purpose was to adopt cluster stars as a set of standards to derive calibration curves for the
observing nights. These will then be applied to calibrate the photometry of bulge clusters observed 
soon after or before the calibrating clusters. 
In order to calibrate the photometry of NGC~6528 we observed 
NGC~6752 during the same nights, collecting 6$u$-, 6$v$-, 6$b$- and 6$y$-band 
images for a total FoV of $\approx$ 4\min$\times$4\min centered on the cluster.
These observations overlap with the FoV of our published photometry for 
NGC~6752, that will be later adopted to calibrate the data set. 
The log of NGC~6752 observations is also shown in Table~1.

 \begin{figure*}
   \includegraphics[width=0.5\textwidth,height=17.5truecm]{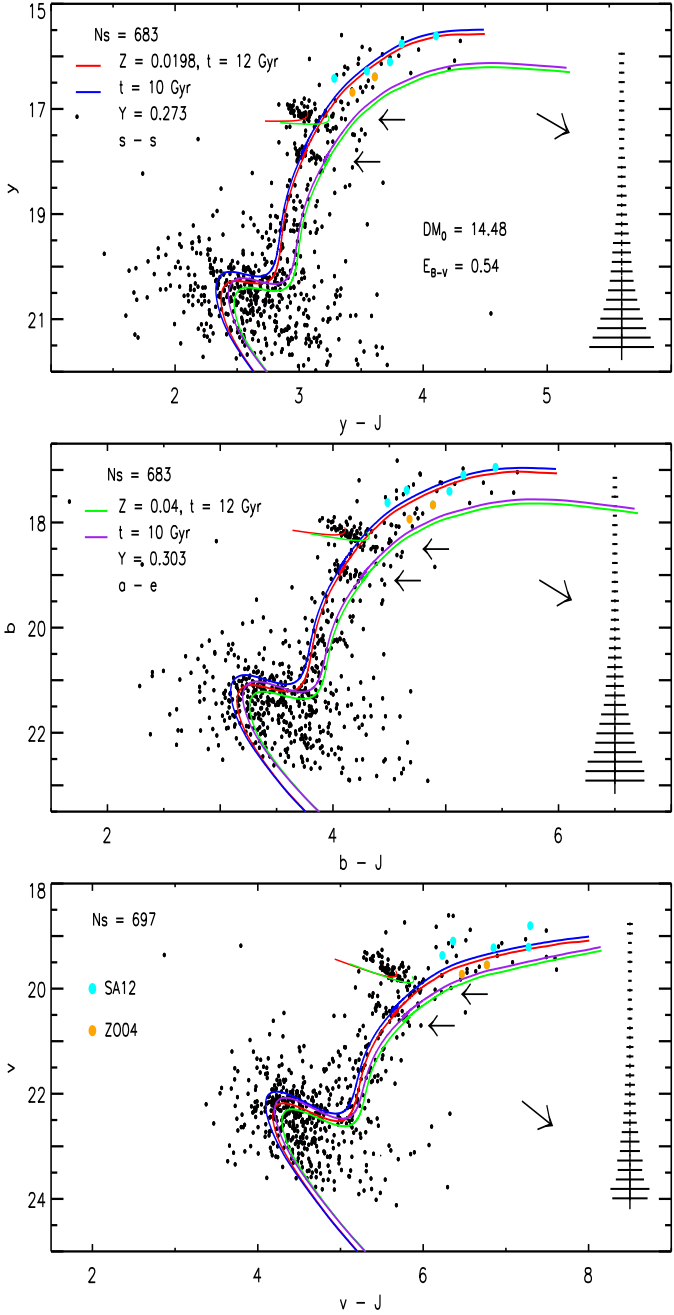}
      \caption{$y, \ y-J$ (top), $b,\  b-J$ (middle), $v,\ v-J$ (bottom) CMDs of NGC~6528. 
      Stars have been selected according to proper motions and photometric accuracy and corrected 
      for differential reddening.
      The red and green lines display cluster isochrones at fixed age but
      different chemical compositions. The blue and the purple lines display 
      younger isochrones for the same chemical compositions (see labeled values).
      The adopted true distance modulus and mean reddening are labeled.
      The cluster isochrones are based on evolutionary models computed by assuming either 
      a scaled-solar (s--s, red and blue) and an $\alpha-$enhanced ($\alpha$--e) 
      chemical mixture (green and purple). 
      Orange and cyan filled dots mark RGs with spectroscopic abundances from ZO04 and SA12, respectively.
      The two horizontal arrows mark the location of the RHB (brighter) and of the RGB bump
      (fainter), while the right bended arrow shows the reddening vector.}
         \label{fig2}
   \end{figure*}
   
Raw images were pre-processed by using tasks available in the IRAF data analysis 
environment for bias subtraction and flat-fielding. To flat-field these data we adopted median 
sky flats collected during the three observing nights. The photometry was performed using 
DAOPHOT$\,${\footnotesize IV}/ALLSTAR and ALLFRAME \citep{ste87, ste91, ste94}. 
We first estimated an analytical point-spread function (PSF) for each frame by selecting bright, 
isolated stars, uniformly distributed on the chip. A Moffat analytical 
function quadratically variable on the chip was assumed for the PSF. 
We performed PSF analytical photometry on each image with the task 
ALLSTAR. In order to obtain a global star catalog for each cluster and each night,  
we used the task DAOMATCH/DAOMASTER \citep{ste94} and then performed simultaneous 
PSF-fitting photometry with the task ALLFRAME.
The final star catalogs of the two nights of observations of NGC~6528
include $\approx 22,000$ (July 11) and  $\approx 10,000$ stars (July 13), with a measure in at 
least two bands. 
   
The NIR photometry of NGC~6528 is from SOFI (Son of ISAAC) on NTT and VIRCAM on 
VISTA (Visible and Infrared Survey Telescope for Astronomy, Paranal, ESO).
The SOFI data set consists of 111 $J$-, 12 $H$- and 111 $K_s$-band images 
(see the green footprints in Fig.~1). The images were pre-reduced with a specific IRAF task 
pipeline kindly provided by M. Dall'Ora. As for EFOSC2, the photometry of the SOFI images 
was carried out with 
DAOPHOT$\,${\footnotesize IV}/ALLSTAR. A single list of improved positions and instrumental magnitudes 
for the stars located in overlapping fields was obtained by using DAOMATCH/DAOMASTER.
   
The VISTA Variables in the V\'ia L\'acte\`a (VVV) Survey  \citep{Minniti2010, Catelan2011} is one of six ESO 
Public Surveys operating on VISTA, scanning the Galactic bulge ($-10\leq l\leq+10$, $-10\leq b\leq +5$) and the 
adjacent part of the southern disk ($-65\leq l\leq-10$, $-2\leq b\leq +2$).
The survey collects data in five NIR bands ($YZJHK_\mathrm{s}$) with the VIRCAM camera \citep{Emerson2010}, 
an array of sixteen 2048$\times$2048~pixel detectors with a pixel scale of 0\farcs 341.
VVV images extend several magnitudes fainter than the Two Micron All Sky Survey \citep[2MASS,][]{2MASS}, 
and enjoy increased spatial resolution \citep{Saito2012}.
Both of these factors are  particularly important for mitigating contaminated photometry in crowded regions such as
the cores of globular clusters.
We retrieved from the Vista Science Archive website\footnote{http://horus.roe.ac.uk/vsa/} VVV images 
containing NGC~6528  for a total FoV of 17\min$\times$22\min (see purple footprint in Fig.~1). 
We also retrieved data for NGC~6624, a cluster adopted later to test the validity of metallicity calibrations.
Data were pre-reduced at the Cambridge Astronomical Survey Unit (CASU)\footnote{http://casu.ast.cam.ac.uk/} 
with the VIRCAM pipeline \citep{Irwin2004}. We then performed PSF-fitting photometry by using the 
VVV-SkZ\_pipeline code \citep{VSpaper} on the single 2048$\times$2048~pixel chips extracted from 
the stacked VVV pawprints \citep{Saito2012}. A quadratically variable Moffat function with $\beta =$ 3.5 
has been adopted in this case to perform the PSF-fitting photometry on the VIRCAM images.

\section{Photometric calibration}

\subsection{EFOSC2 data set}

We first applied aperture corrections to the single cluster instrumental photometric
catalogs. The corrections were estimated for the adopted reference $y,b,v,u$-band images 
and then applied to the mean magnitudes. Photometry was also corrected according to 
the exposure times and airmass values of the reference images by adopting the standard 
La Silla extinction coefficients for the \strom filters.

The two corrected instrumental catalogs of NGC~6752 for the nights July 11 and 13 
were cross-correlated with the calibrated \strom catalog for this cluster. We ended up with a sample
of $\approx 5,000$ and $\approx 1,500$ stars in common for the two nights, respectively, 
to estimate the calibration curves. The stars were selected
in photometric accuracy, $\sigma_{y,b,v,u} <$ 0.1 mag, and the cluster center
was excluded by keeping stars with a distance from the cluster center larger than 1.5\min.
The derived calibration curves for the two nights are:\\

July 11:\\

$u= u_i - 1.911(\pm0.022) + 0.127(\pm0.010)\times(u_i-b_i)$,\\

$v= v_i - 0.907(\pm0.005)$,\\

$b= b_i - 1.215(\pm0.005) + 0.055(\pm0.0011)\times(b_i-y_i)$,\\

$y= y_i - 1.347(\pm0.004) + 0.070(\pm0.009)\times(b_i-y_i)$\\

July 13:\\

$u= u_i - 1.966(\pm0.012) + 0.063(\pm0.006)\times(u_i-b_i)$,\\

$v= v_i - 0.993(\pm0.005)$,\\

$b= b_i - 1.240(\pm0.005) + 0.034(\pm0.0012)\times(b_i-y_i)$,\\

$y= y_i - 1.388(\pm0.005) + 0.043(\pm0.012)\times(b_i-y_i)$\\

where $i$ stands for $instrumental$ magnitude.

We applied these calibration curves to NGC~6528 corrected instrumental 
photometry for the nights of July 11 and 13.  We then merged the two data sets 
obtaining a final calibrated catalog of $22,986$ stars.

The typical accuracy of the absolute zero-point calibration is 
$\sim$ 0.06 mag for the $u$-band data and $\sim$ 0.05 mag for 
the $v,b, y$-band data.

 \begin{figure}
   \includegraphics[width=0.47\textwidth,height=9truecm]{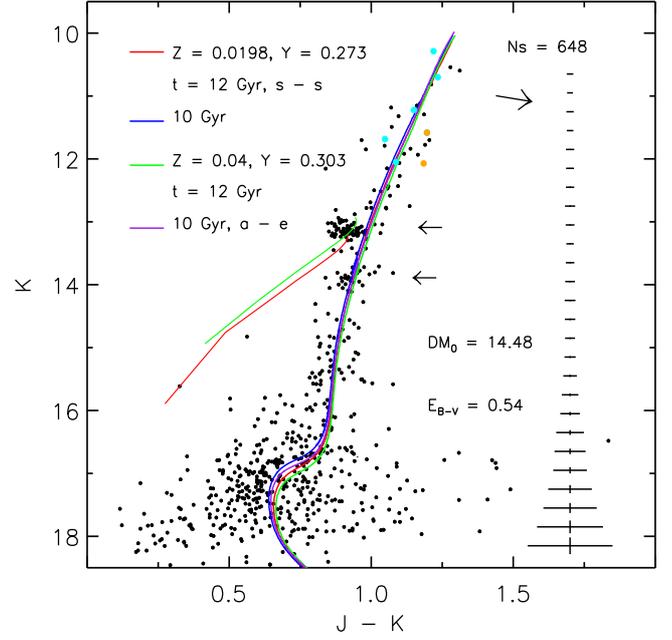} 
      \caption{$K, \ J - K$ CMD of NGC~6528. Stars have been selected
      according to proper motions and photometric accuracy and corrected for differential 
      reddening. Symbols and lines are the same as in Fig.~2.}
	 \label{fig3}
   \end{figure}
   
\subsection{SOFI and VIRCAM data sets}
The NIR broad-band filters $J, H, K_S$, with which the SOFI images were collected, 
have passbands very similar to those used for the observations of the $2$MASS survey \citep{cutri}. 
Therefore, after the aperture corrections were computed for the PSF stars in each single frame 
by using DAOGROW \citep{ste90}, the instrumental magnitudes were calibrated to the 2MASS
system using a set of local standards. The solutions of the calibration
equations were found with CCDSTD \citep{ste93}. 
We ended up with a catalog of $\sim$ 65,000 stars, with at least one measurement in two bands. 
The photometric catalog ranges from the tip of the RGB down to $\sim$ 0.5 - 1 mag below the Main 
Sequence Turn-Off (MSTO).

The VIRCAM photometric catalog was also tied to the 2MASS system. The calibration consisted 
in applying the classical correction for the zero point and color terms to the ALLFRAME output. 
The procedure has already been described by \citet{MoniBidin2011, Chene2012,VSpaper}.
The use of the VVV-SkZ\_pipeline was fundamental for this work, since it is the only photometric 
procedure for VVV data that manages to provide accurate photometry even for partially saturated stars.
The $H_{2MASS}$-band photometry (hereafter $H_{2MASS} = H, J_{2MASS} = J, Ks_{2MASS} = K$)
was neglected, since the precision in this filter was not good enough at the limiting magnitudes we are dealing with.  

The SOFI and VIRCAM catalogs were matched and we estimated the weighted mean
of $J$ and $K$-band magnitudes. 
The final merged NIR catalog includes $\approx 50,000$ stars, with an
accuracy of  $\approx$ 0.05 mag at $J \approx$ 19 mag, and covering a FoV of $\approx$ 8\min$\times$8\min.
\section{\stromc--NIR photometry}
\subsection{NGC~6528}

The \strom photometry of NGC~6528 was cross-correlated with the NIR photometry 
and we ended up with a \stromc--NIR catalog of $22,025$ stars. 
This catalog was then matched to
optical and NIR photometry collected with ACS@HST and WFC3@HST 
(Program IDs: GO9453, GO11664). The reader is referred to LA14 for 
details on this data set observations and reduction techniques.
The final merged catalog includes $3,189$ stars and covers a 
FoV of $\approx$ 2.2\min$\times$2.2\min,
with the cluster center located in the North-East corner (see red footprint in Fig.~1).
We corrected the photometry for differential extinction by adopting the individual 
reddening values estimated for each star in the field by LA14. 

We then adopted the proper-motion selection of  LA14 to obtain a clean 
sample of $737$ cluster stars with a measurement in at least two
\strom and two NIR bands. The matched catalog reaches 1 mag below the turn-off (TO)
with an accuracy of $\approx$ 0.2 mag at $y \approx$ 21.5 mag.

We also matched current NGC~6528 photometry with the high-resolution 
spectroscopic targets by \citet[hereafter ZO04, ZO08]{zocc04} 
and \citet{origlia}, and $CaT$ spectroscopy by \citet[hereafter SA12]{saviane}.
The three RGs in the ZO04 sample are located inside our FoV, 
namely OGLE~357459, OGLE~357480 and I~42, while only one 
out of the four RGs in the Origlia et al. sample (star 3167), is covered by our catalog.
Star OGLE~357480 has a color systematically bluer than cluster 
RGs and it does not belong to NGC~6528 according to our proper-motion selection;
the same outcome applies to star 3167. 
We ignored both of them. Six stars of SA12 sample were found 
in common with our \stromc--NIR photometry. One star, R1-42, 
is star I~42 of ZO04 sample, and the other five, R2-8, R2-41, 1\_3704, 1\_1044 and 1\_2735, 
belong to NGC~6528 according to current proper-motion selection. 
We ended up with seven candidate cluster member RGs with 
spectroscopic abundance measurements available.

  \begin{figure}
   \includegraphics[width=0.5\textwidth,height=17.5truecm]{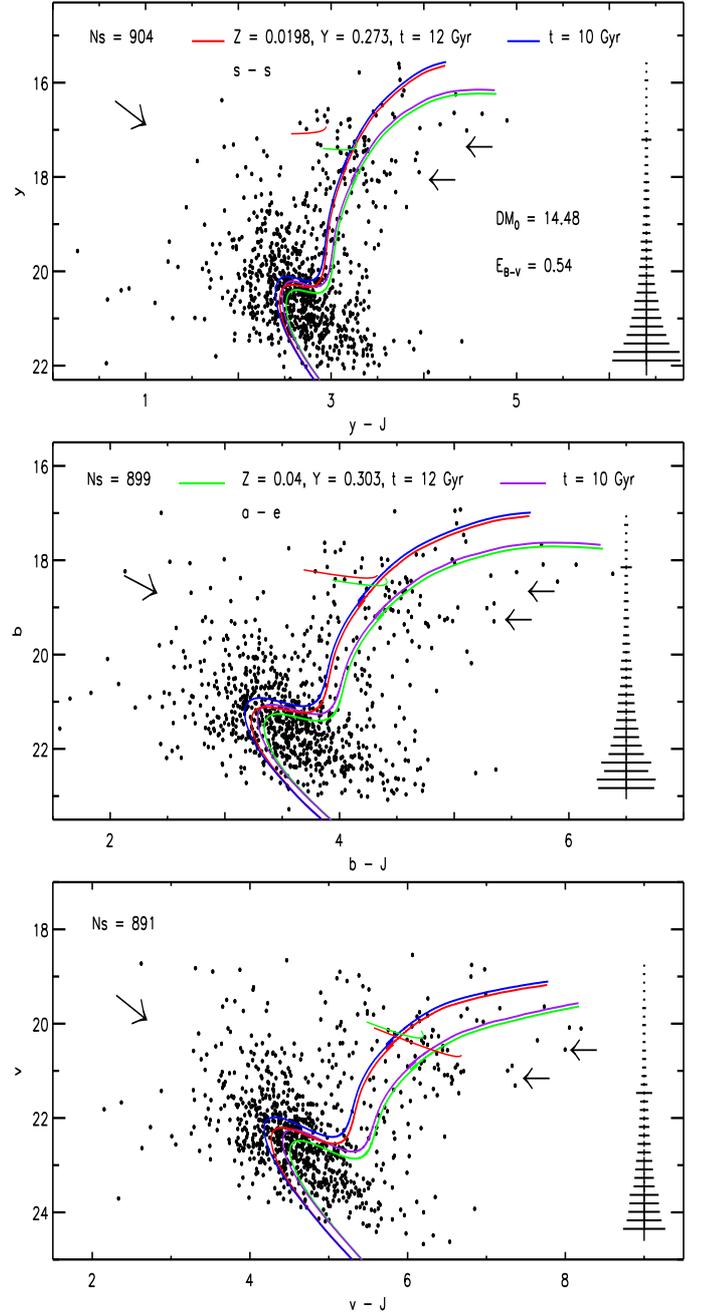} 
      \caption{Same as Fig.~2 but for the field surrounding NGC~6528.
      Symbols and lines are the same as in Fig.~2.}
	 \label{fig4}
   \end{figure}

Fig.~2 shows the $y, \ y-J$,  $b,\  b-J$, and $v,\ v-J$ proper-motion cleaned color-magnitude diagram (CMD) 
of NGC~6528. Stars have been selected according to the ''separation index''\footnote{The ''separation index'' 
quantifies the degree of crowding of a star after PSF photometry has been performed (Stetson et al.\ 2003).}, 
by keeping the best 90\% of the stars for each 0.18 magnitude bin in the ranges 15.5 $< y <$ 22, 
16.7 $< b <$ 23.1 and 18.5 $< v <$ 24.3 mag. 
We excluded the $u$-band since the photometry is based on only one image per night and 
it is not accurate enough for this study.

In this analysis, we adopt three different CMDs since, from the empirical 
point of view, different \stromc--NIR bands have different 
photometric accuracy. Moreover, from the theoretical point of view,
the Color--Temperature Relations (CTRs) adopted to transform isochrones 
into the observational plane need to be tested.
This is the first time, indeed, that these models are adopted to fit the 
entire magnitude range from the MS to the RGB in \stromc--NIR CMDs.

The solid lines in Fig.~2 show isochrones for different ages and different chemical composition, 
namely  $Z$ = 0.0198, $Y$ = 0.273, scaled-solar, $t$ = 12 (red solid line) and $t$ = 10 Gyr (blue),
and $Z$ = 0.04, $Y$ = 0.303, $\alpha$-enhanced (\afe = 0.4), $t$ = 12 (green) and $t$ = 10 Gyr (purple).
The corresponding scaled-solar and $\alpha$-enhanced Zero Age Horizontal Branches (ZAHBs, red and green
solid lines) for $Z$ = 0.0198 and $Z$ = 0.04 are also plotted.
The adopted chemical compositions have the same iron content, \feh $\sim$ 0.06, but different helium
and $\alpha$-element abundances.
Isochrones and ZAHBs are from the BASTI data base \citep[hereafter PI06]{pietri04,pietri06} 
and evolutionary prescriptions were transformed into the observational plane by using atmosphere 
models computed assuming scaled-solar and $\alpha$-enhanced chemical mixtures. 
We adopt a true distance modulus of $DM_0 = 14.50$ mag (LA14), 
and a mean reddening of $E(B-V) = 0.54$ mag for NGC~6528. The reddening was
estimated as a weighted mean of the values given by the extinction map
of \citet{oscar} for this FoV and by adopting a standard reddening law,
$R_V = A_V/E(B-V) = 3.1$. The extinction coefficients are then estimated by applying 
the \citet{card89} reddening relations, finding 
$A_{y} = A_V$,  $A_{b} = 1.225 \times A_V$, $A_{v} = 1.424 \times A_V$, and 
$E(y - J)= 2.23 \times E(B-V)$ mag, $E(b - J)= 2.91 \times E(B-V)$ mag, and
$E(v - J)= 3.527 \times E(B-V)$ mag.

\begin{figure}
\includegraphics[width=0.47\textwidth,height=9truecm]{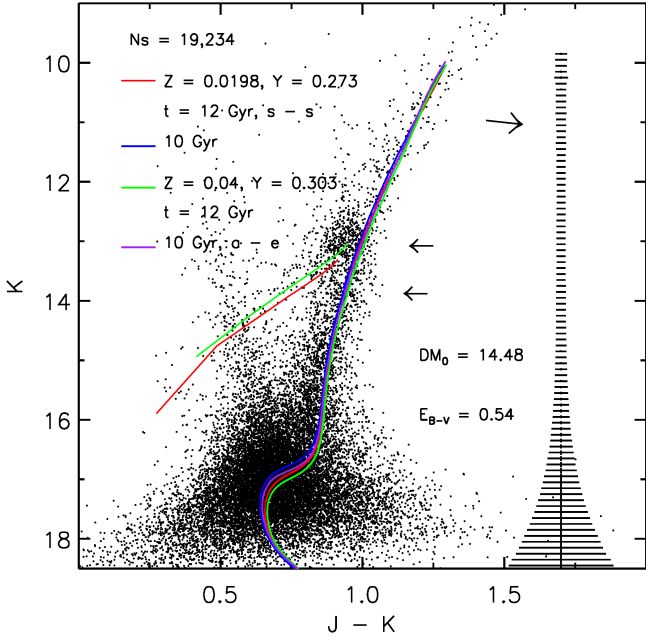}  
\caption{Same as Fig.~3, but for the field surrounding NGC~6528.
Stars have been selected according to photometric accuracy. Cluster isochrones and ZAHBs are 
plotted by adopting the same distance modulus and reddening. From top to bottom the arrows 
mark the direction of the reddening vector, the location of the RHB-RC and the RGB bump.}
\label{fig5}
\end{figure}

Data plotted in Fig.~2 show a clean sample of NGC~6528 member stars corrected for differential reddening 
down to ~ 1 mag below the TO.  
Unfortunately the precision of both \strom and NIR photometry
(see the error bars in Fig.~2), and the residual uncorrected reddening, 
do not allow us to properly identify the cluster MSTO. 
However, evolved evolutionary features 
can be easily identified in the three CMDs. The RGB bump is identified 
as an over-density at $y \sim$ 17.8 and $y - J \sim$ 3 mag, 
$b \sim$ 17.8 and $b - J \sim$ 4 mag and $v \sim$ 20.5 and 
$v - J \sim$ 5.7 mag (see the arrows in Fig.~2). The RHB is well separated from the RGB and is 
located at $y \sim$ 17 and $y - J \sim$ 3, $b \sim$ 18 and 
$b - J \sim$ 4 mag and $v \sim$ 19.5 and $v - J \sim$ 5.5 mag (see the arrows).
The scaled-solar models (red and blue solid lines) fit 
quite well the observations, within the uncertainties, and suggest for this 
cluster an age of 11$\pm$1 Gyr. 
Interestingly enough, stars from the spectroscopic samples agree quite 
well with scaled-solar isochrones in the three CMDs.
The $\alpha$-enhanced isochrones (green and purple) attain, as expected, 
systematically redder colors in the three CMDs. However, the difference 
in color and in magnitude between the two different sets does not 
allow us to reach firm conclusions concerning the cluster 
$\alpha$-enhancement. 
Although, the observed RGB-bump stars appear to be systematically 
bluer than predicted by the $\alpha$-enhanced isochrones.  

We plotted the same candidate cluster stars in the NIR $K,\ J-K$ CMD in Fig.~3.
In this plane the RHB and the RGB-bump are much better defined (see arrows). 
The same isochrones and ZAHBs of Fig.~2 are over-plotted 
by adopting the same reddening and distance modulus. 
It is worth noticing how the isochrones overlap along the MS and the RGB phases in the $K,\ J-K$ CMD, 
while, at fixed iron abundance, the $\alpha$-enhanced models are redder than the 
scaled-solar models in the \stromc--NIR CMDs (see Fig.~2).
Fig.~3 shows that $\alpha$-enhanced and scaled-solar isochrones agree 
quite well, within the uncertainties (see error bars in the figure), 
with NIR observations, confirming an age of 11$\pm$1 Gyr for the cluster.
This result is in very good agreement with the age derived by \citet{feltz02}
for NGC~6528, i.e. $11\pm2$ Gyr. They adopted $F555W, F814W$ WFPC2 
photometry and astrometry to obtain a proper-motion cleaned 
CMD for the cluster. To fit the optical CMD they use $\alpha$-enhanced isochrones
by \citet{salasnich} for $Z =$0.04, a distance modulus of $DM_0 =$ 14.29 mag
and reddening $E(B-V)=$ 0.54 mag. 
Moreover, current estimates are also in fairly good agreement with 
the results of \citet{momany03}.
They presented NIR photometry of NGC~6528 based on SOFI data - that are included in our data set -  
and cluster members  were selected by adopting HST proper motions. They estimated 
a cluster age of $12.6$ Gyr, by adopting \citet{bertelli94} scaled-solar isochrones 
for $Z = 0.02$ and a distance modulus $DM_0 = 14.44$ mag and $E(B-V) = 0.55$ mag.
Finally, our age estimate is in very good agreement with the results of LA14
that find $t =$ 11$\pm1$ Gyr, by fitting their optical HST CMD with 
a scaled-solar isochrone with \feh = 0.20 \citep{pisa}, and by adopting
$DM_0 = 14.50$ mag and $E(B-V) = 0.56$ mag.

\begin{figure}
\includegraphics[width=0.5\textwidth,height=0.5\textwidth]{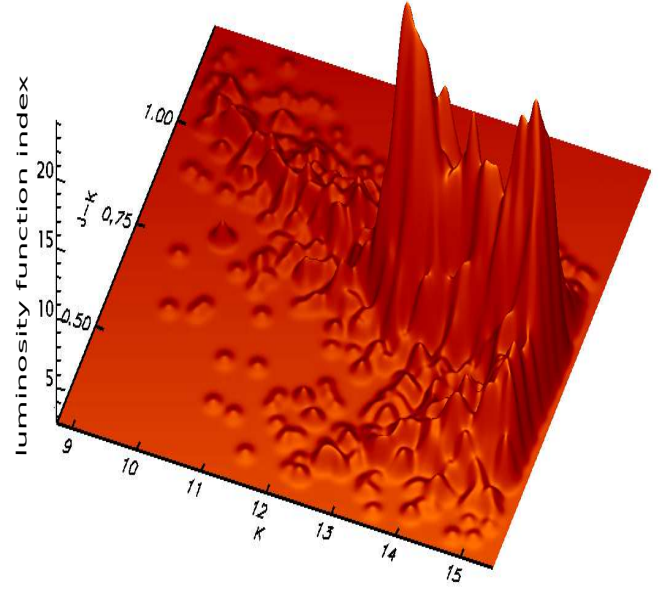}  
\caption{
3D $K,\ J-K$ CMD of field stars surrounding NGC~6528, 
based on VIRCAM-SOFI photometry. The CMD is slightly rotated to 
better show the separation between blue MS and RG stars.
At $K =$ 14, MS stars cluster around $J-K =$ 0.6, 
while RGs around $J-K =$ 0.8 mag.}
\label{fig6}
\end{figure}

\subsection{The field surrounding NGC~6528}
A clean proper-motion sample of 986 candidate bulge stars is selected from 
the \stromc--NIR catalog
and plotted on the $y, \ y-J$,  $b,\  b-J$, and $v,\ v-J$ CMDs in Fig.~4.
In the comparison between theory and observations we adopted the same 
theoretical framework, reddening and absolute distance modulus as for NGC~6528.
Fig.~4 shows that we do not have enough statistics to properly characterize the 
bulge stellar populations in the Baade's window. However, data plotted in the  
the \stromc--NIR CMDs bring forward a few interesting features:

{\em a)} The RGB shows a clear dispersion in color in all the CMDs, 
mostly given by the presence of both a metallicity and an age spread.
Indeed, the adopted isochrones do not bracket the color dispersion along 
the RGB. Together with the above intrinsic properties the color dispersion 
is also caused by the effect of residual uncorrected differential reddening 
(see the arrows in Fig.~4 for the reddening direction), by depth effects 
and by photometric errors. Current data set does not allow us to disentangle 
the above degeneracies.

{\em b)} The RHB is not a well defined sequence and red clump (RC) stars are overlapping the RGB.

{\em c)} The RGB-bump is not clearly visible, except for a small overdensity of stars ranging
from 19.5 $\lesssim v \lesssim$ 21 mag and 5 $\lesssim v - J \lesssim$ 6.5 mag
in the $v,\ v - J$ CMD.

{\em d)} There is a small residual contamination by disk stars located in a sequence 
brighter and bluer than the MSTO in all the CMDs.

Data plotted in Fig.~4 show that a fraction of observed stars 
along the MS and the RGB are systematically redder and/or fainter 
than predicted by adopted cluster isochrones. This result would suggest 
the possible occurrence of large samples of super metal--rich stars, \feh $>$ 0.1, in the Baade's window.
However, these objects could also be explained with an increase either 
in the mean reddening or in depth or both.  

\subsection{The RC-RHB stars and the RGB-bump}
To shed new light in this interesting open problem, we plotted the 
entire selected VIRCAM-SOFI catalog in the $K,\ J-K$ CMD in Fig.~5. 
The RC-RHB region and the RGB-bump are much better defined compared
to Fig.~4 due to the increased statistics and to the 
reduced sensitivity of NIR colors to effective temperature when 
compared to optical--NIR colors.
The same isochrones of Fig.~4 are over-plotted
by adopting the same reddening and distance modulus. 
This figure shows that the adopted isochrones also agree well with the
NIR observations of the bulge stars in the Baade's window. In particular, 
the dispersion in color of the bulge MS is quite similar to the 
dispersion in color of the NGC~6528 CMD at the same magnitude 
level (see Fig.~3). This suggests that the spread in color is 
mainly due to photometric errors and residual differential reddening.

\begin{figure}
\begin{minipage}[l]{0.8\textwidth}
\centering
\hspace{-5.5truecm}
 \includegraphics[width=0.55\textwidth,height=0.28\textheight,angle=0]{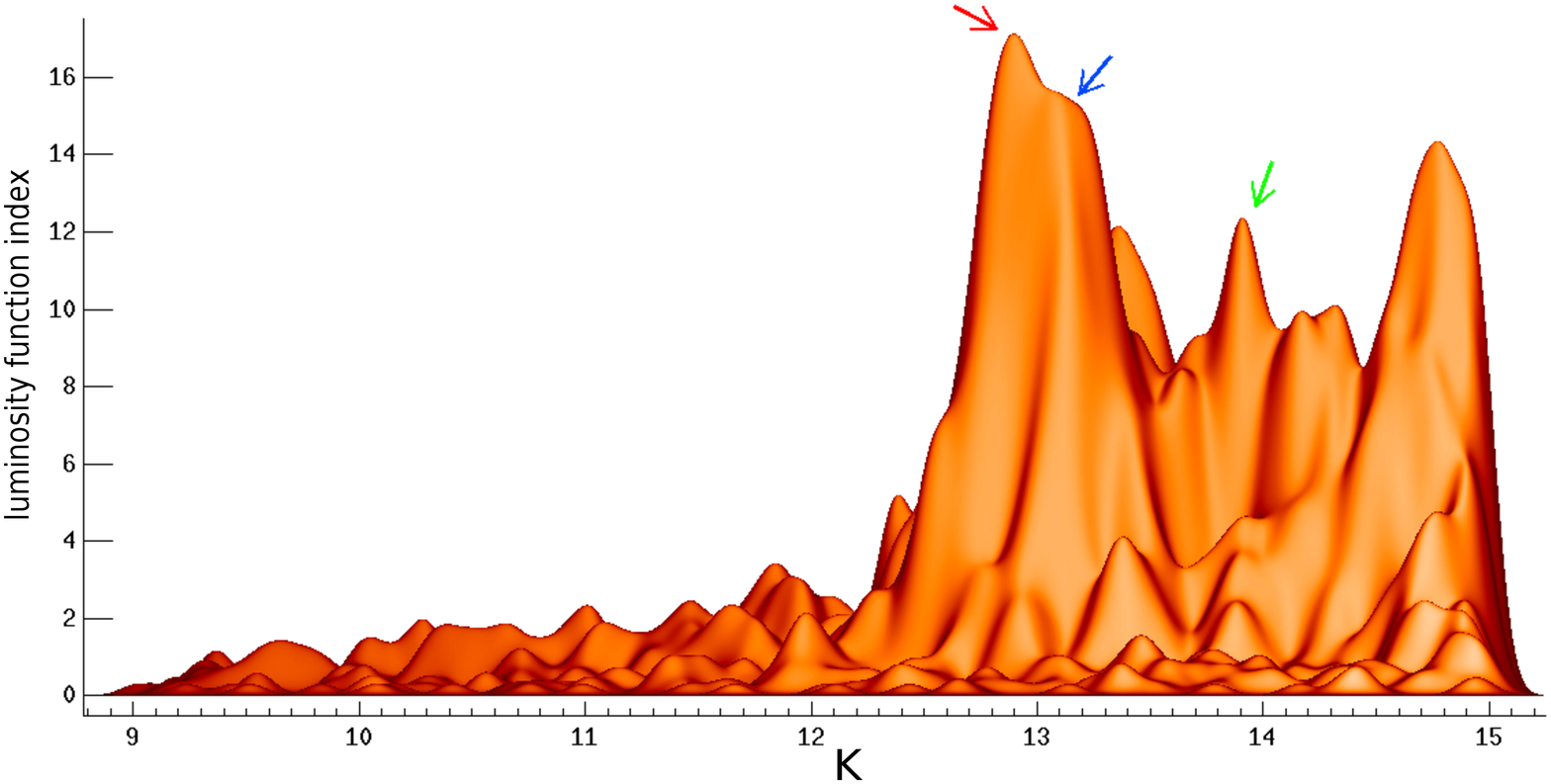}    
     \end{minipage}
 \begin{minipage}[r]{0.5\textwidth}
\centering
 \includegraphics[width=1\textwidth,height=0.3\textheight,angle=0]{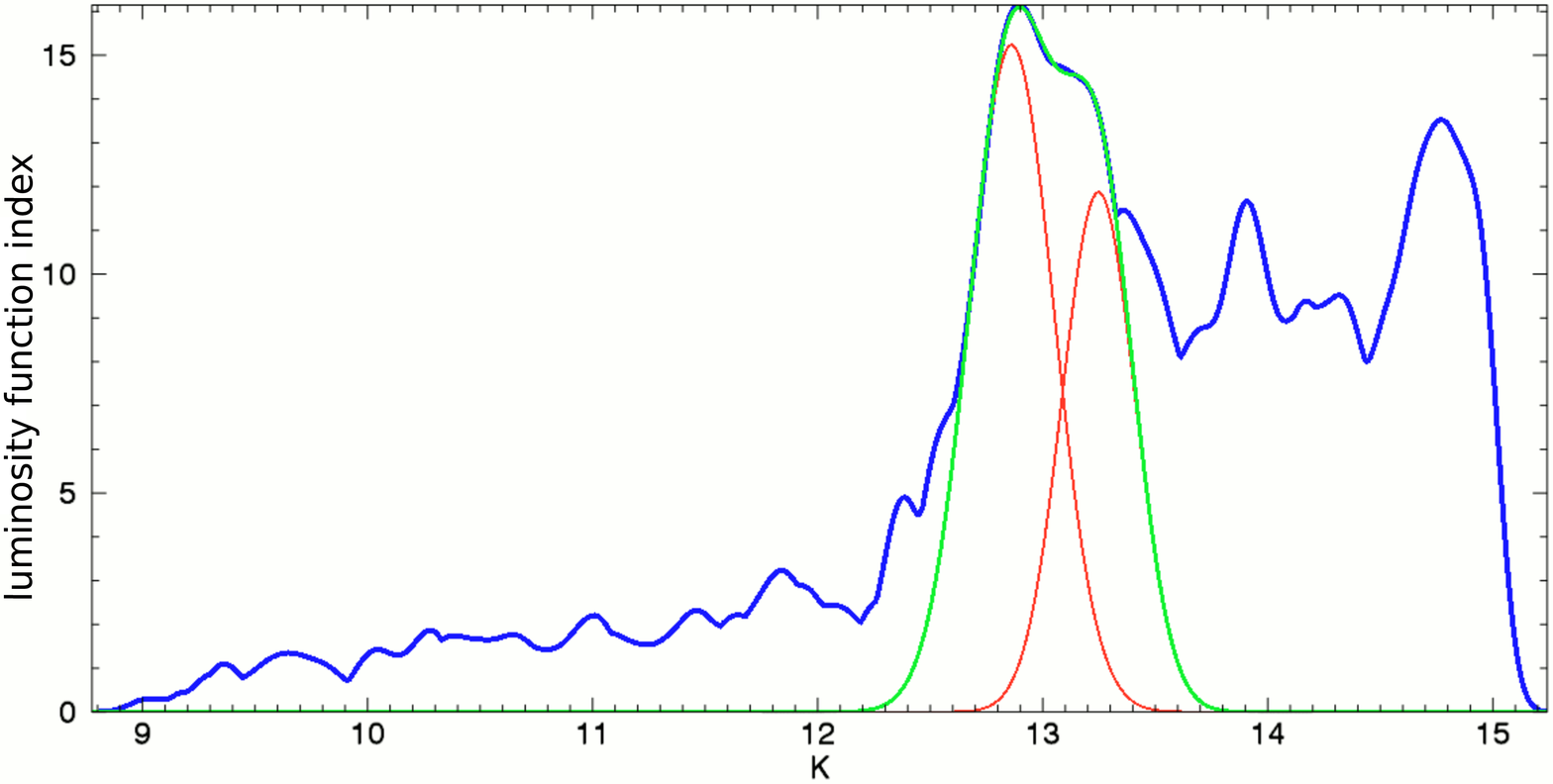}
\centering
\vspace{-0.2truecm}
\includegraphics[width=1\textwidth,height=0.3\textheight,angle=0]{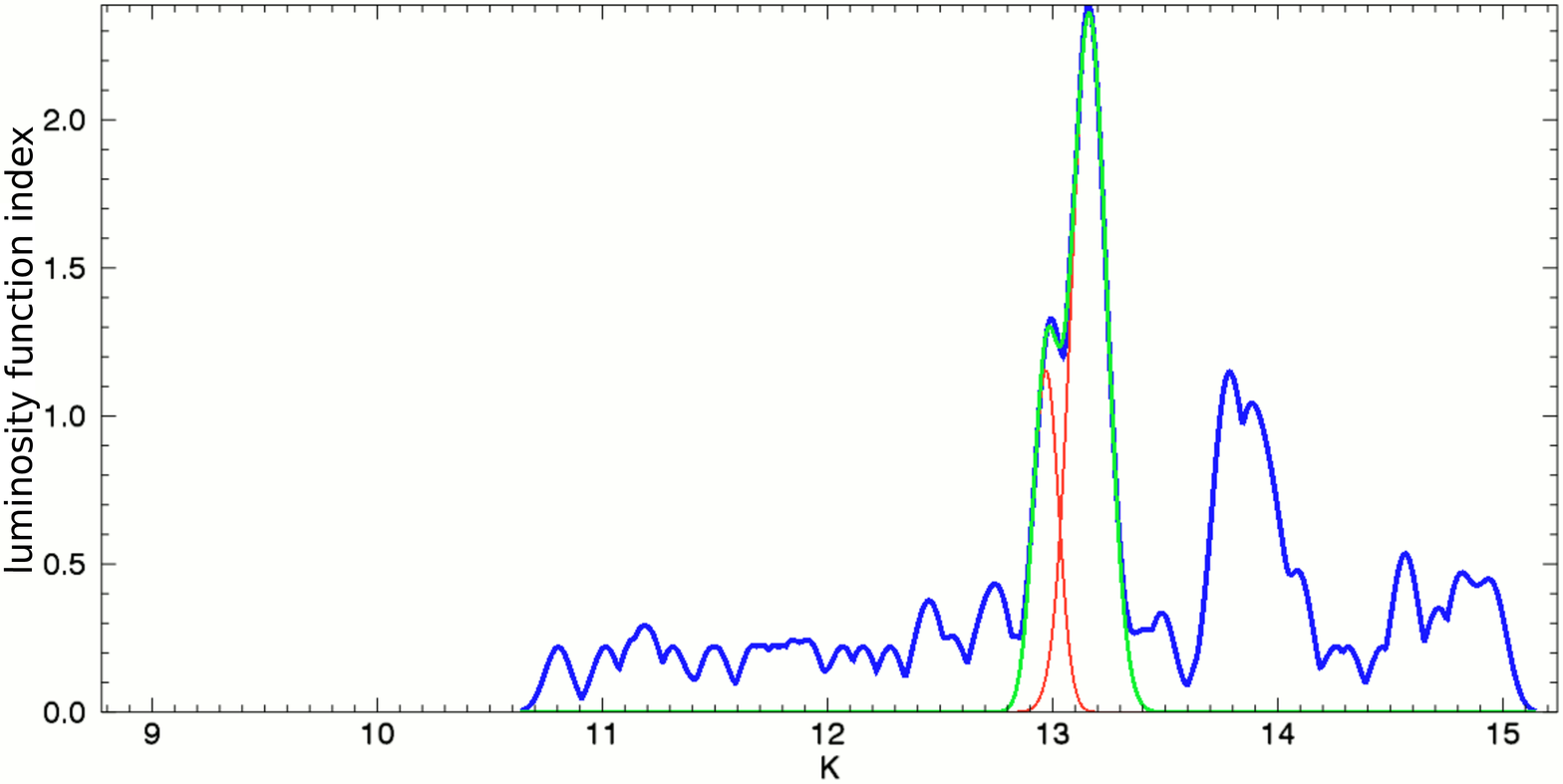}
\caption{Top: 2D $K$-band -- based on VIRCAM-SOFI photometry -- 
luminosity function of field RGs surrounding NGC~6528.
Middle: Projected $K$-band luminosity function of field RGs. The two Gaussian functions adopted 
to fit the RC-RHB regions are over-plotted as red solid lines. The green line 
shows the cumulative fit. 
Bottom: Projected $K$-band luminosity function of NGC~6528 RGs selected according 
to proper motions. Symbols and lines are the same as in the middle panel.}
 \end{minipage}     
         \label{fig7}
   \end{figure}
  
However, data plotted in this figure show that the ZAHBs attain 
a $K$-band magnitude that is systematically fainter than observed RHB-RC stars.
This is an interesting empirical evidence. Indeed, current spectroscopic 
targets use as stellar tracers RC stars. The selection is 
performed using NIR CMDs and the region in color and in magnitude 
they adopt includes not only truly RHB (old [$t >$10 Gyr], 
low-mass), but also truly RC (intermediate-age [$t <$ 8 Gyr], 
intermediate-mass) stars. Note that the above separation is 
far from being an academic debate and, indeed, according to 
current evolutionary prescriptions, in the former group the 
central helium-burning takes place in a more massive, electron-degenerate 
core, while in the latter takes place in a less massive, partially 
electron-degenerate core \citep{fiorentino, coppola}.  

The above scenario does imply a difference in mean magnitude between 
the two groups, since RHB stars are in optical and in NIR magnitudes  
fainter and bluer than RC stars. To further constrain this effect 
we adopted the entire sample of VIRCAM-SOFI measurements to study the magnitude and
color distribution of stars in the RHB-RC region. 
To avoid NGC~6528 we selected only stars with a distance from the cluster 
center larger than $\approx$ 2.5\arcmin.
Fig.~6 shows the 3D $K,\ J-K$ CMD of candidate stars brighter than
$K =$ 15.5 mag. To select only RGs we then performed a color selection 
by excluding all stars bluer than $J-K =$ 0.7 mag.
Fig.~7 (top panel) shows the 2D luminosity function of selected RGs.
Central helium burning stars (RC, RHB) display a double peak in the magnitude 
range 12.5 $<K<$ 13.5 mag (red and blue arrows, respectively), while
the RGB-bump shows a well defined peak at $K =$ 13.90$\pm$0.05 mag (green arrow).
The middle panel of the same figure shows the projected $K$-band
luminosity function of selected RG stars (blue solid line). 
We fit the RC-RHB region
by using two Gaussian functions (red lines), finding the following peaks:
$K =$ 12.86$\pm$0.02 mag ($\sigma$ = 0.19 mag),
and $K =$ 13.25$\pm$0.02 mag ($\sigma$ = 0.16 mag).
It is noteworthy that the fainter peak is also systematically bluer ($J-K \approx$ 0.9 mag) 
than the brighter one ($J-K \approx$ 1.0 mag, see Fig.~6). 
This suggests a clear separation both 
in magnitude and in color between the two peaks.

The bottom panel of Fig.~7 shows the same projected $K$-band luminosity function
of bright ($K \gtrsim$ 15.5 mag) proper-motion selected RGs in NGC~6528.
The RHB peaks at $K \approx$ 13.15 mag, while the RGB-bump at $K =$ 13.85$\pm$0.05 mag.
The RGB-bump magnitude is in very good agreement with the bump luminosity 
found by \citet{momany03}, $K =$ 13.85$\pm$0.05 mag, 
and by \citet{ferraro00},  $K \approx$ 13.80 mag.
We then performed a fit of the RHB region by adopting two Gaussian functions (red lines)
with peaks at $K =$ 13.16$\pm$0.02 mag ($\sigma$ = 0.08 mag) and $K =$ 12.97$\pm$0.02 mag 
($\sigma$ = 0.06). The presence of the secondary brighter peak among cluster RGs 
might be due to minor residual contamination by field stars. 
Moreover, the faintest peak at $K =$ 13.16$\pm$0.02 mag is in very good agreement with 
$K =$ 13.20$\pm$0.05 mag found by \citet{momany03} for the RHB of NGC~6528.

The above findings indicate that the fainter peak of field RG luminosity function at $K =$ 13.25 mag
might be associated with the RHB of the old stellar population, while the 
brighter peak at $K =12.86$ mag to RC stars.
Interestingly enough, the difference in magnitude between the RC and the RHB peak 
($\Delta K \sim$ 0.4 mag) supports recent findings concerning the boxy/peanuts shape of the Galactic 
bulge (McWilliam \& Zoccali 2010; Saito et al. 2011).

\section{Calibration of new visual--NIR metallicity indices for giant stars}

The calibration of \strom photometric indices to obtain stellar metal 
abundances is not a new technique. Empirical calibrations based on such 
a method have been given by \citet{strom64, bond70, craw75, nissen81}. 

All the derived relations to estimate the metal abundance of RG stars are hampered 
by the presence of molecular $CN$,$CH$ and $NH$-bands that affect the \strom $uvb$ 
filters, and in turn the global metallicity estimates. Moreover, most of the relations presented
in literature include the $c_1$ index, that is based on the $u$ filter. 
Observations in the $u$ filter are very demanding concerning 
the telescope time and the photometry in this band is less accurate due to the 
reduced CCD sensitivity in this wavelength region. For more information about 
the \strom photometric system and metallicity calibration see \citet{io07, io12}.

  \begin{figure*}
   \includegraphics[width=17truecm, height=17truecm]{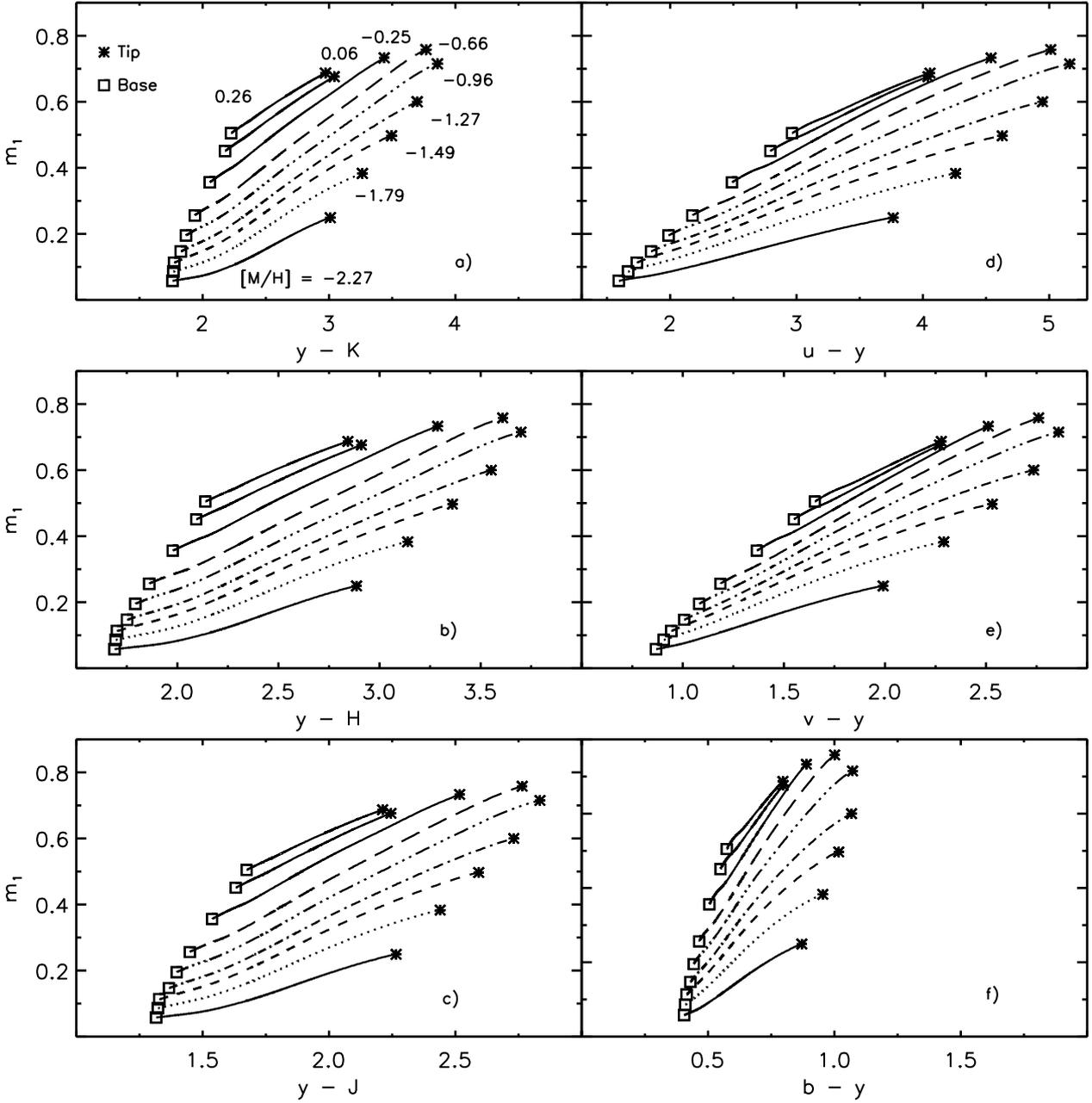}
      \caption{Left panels -- $m_1$ vs $y-K$ plane [panel a)] for 
         isochrones at fixed cluster age ($t =$12 Gyr) and different global 
         metallicities ($\mh$, see labeled values). The evolutionary phases 
         range from $\approx$ the base (empty squares) 
         to $\approx$ the tip of the RGB (asterisks). Evolutionary tracks were computed 
         by assuming a scaled-solar chemical mixture and transformed 
	 into the observational plane by adopting atmosphere models computed 
         assuming the same mixture. The panels b) and c) show similar 
         relations, but in the $m_1$ vs $y-H$ and in the $m_1$ vs $y-J$ plane. 
         Right panels -- Same as the left, but for the $m_1$ vs $u-y$ 
         [panel d)], $m_1$ vs $v-y$ [panel e)], $m_1$ vs $b-y$ [panel f)] 
         planes.}
	 \label{fig7}
   \end{figure*}
   
In \citet{io12} we derived, for the first time, a theoretical calibration of a metallicity diagnostic 
based on the $m_1$ index and on visual--NIR colors for dwarf stars. 
We provide in this paper a similar theoretical calibration based on the same colors but for RG stars.

Independent MIC relations were derived using cluster isochrones based 
on both scaled-solar and $\alpha$-enhanced ($\afe=0.4$) evolutionary models (PI06).
Theoretical predictions were transformed into the observational plane
by adopting bolometric corrections (BCs) and CTRs based on atmosphere models computed assuming the same heavy 
element abundances \citep[PI06,][]{CK06}. 
The Vega flux adopted is from \citet{CK94}\footnote{The complete set of BCs, CTRs 
and the Vega flux are available at http://wwwuser.oat.ts.astro.it/castelli}.
The metallicities used for the calibration of the MIC relations are: 
Z = 0.0001, 0.0003, 0.0006, 0.001, 0.002, 0.004, 0.01, 0.02 and 0.03. 
The adopted Z values indicate the global abundance of heavy elements in the             
chemical mixture, with a solar metal abundance of ${(\zx)_\odot}=0.0245$. 
The corresponding iron content for the $\alpha$-enhanced models can be obtained by 
using the relationship given by \citet{salaris}: \feh = \mh $- \log(0.638f + 0.362)$, where $\log(f)$ = \afe.

Fig.~8 shows nine scaled-solar isochrones plotted in different visual--NIR (left) 
and \strom (right) MIC planes. The $J,H,K$ filters where transformed into the 2MASS 
photometric system by applying the color transformations by \citet{carpenter}.
Panels a), b) and c) display the $m_1$ index versus 
three visual--NIR colors ($y - K$, $y - H$, $y - J$), while the panels d), e) 
and f) the $m_1$ index versus three \strom colors ($u - y$, $v - y$, $b - y$). 
The evolutionary phases plotted in this figure range from  approximately 
the base (open square) to the tip of the RGB (asterisk).

The two different sets of MIC relations cover similar $m_1$
values but the visual--NIR colors -- panels a),b),c) -- show a stronger 
sensitivity in the faint magnitude limit and an almost linear change 
when moving from metal-poor to metal-rich stellar structures. 
The \strom colors -- panels d),e),f) -- show a minimal sensitivity 
for stellar structures more metal-rich than \mh $\gtrsim -0.25$.  
Moreover and even more importantly, the slopes of the MIC relations 
based on visual--NIR colors are on average shallower than the 
MIC relations based on \strom colors. This means that the former 
indices have, at fixed $m_1$ value, a stronger temperature sensitivity.    
We performed the same comparison with the $\alpha$-enhanced models
obtaining very similar results.

   \begin{figure*}
   \includegraphics[width=0.5\textwidth]{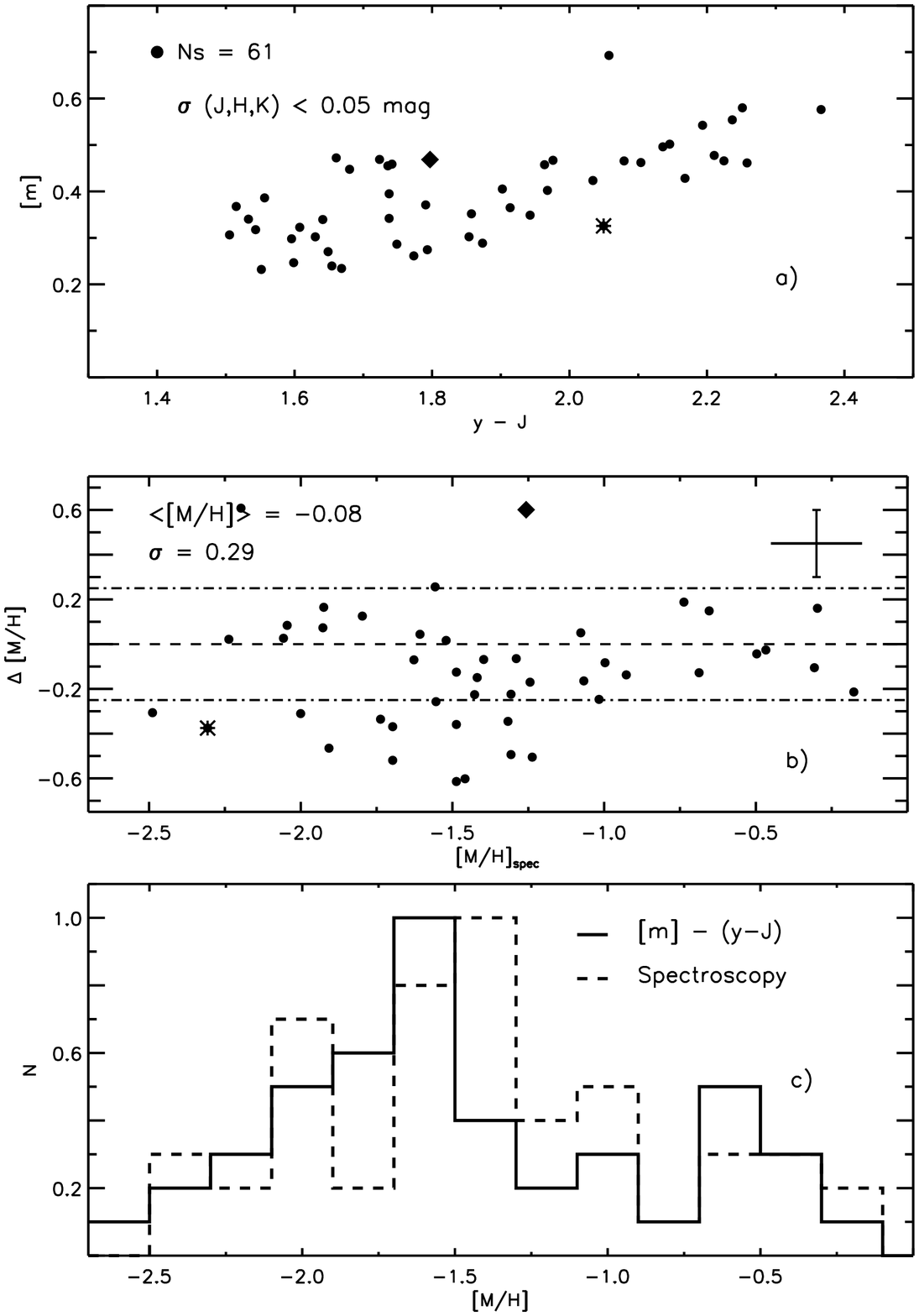} 
      \caption{Panel a) Selected field RGs from the sample of ATT94/ATT98 
       plotted in the $[m],\ y - J$  plane (Ns = 61, filled dots). 
       Panel b) Difference between photometric and spectroscopic metallicities,
       $\Delta \mh = \mh_{\hbox{\footnotesize phot}} - \mh_{\hbox{\footnotesize spec}}$, 
       plotted versus
       $\mh_{\hbox{\footnotesize spec}}$ for the 61 field RGs (filled dots).
       Photometric metallicities are based on the $\alpha$-enhanced $[m], \, y - J$ MIC relation.
       Panel c) Normalized photometric metallicity distribution for the 61 RGs obtained with
       the $[m],\ y - J$ MIC relation (black solid line), compared to the normalized spectroscopic
       distribution (black dashed).}
	 \label{fig8}
   \end{figure*}
   
MIC relations for RG stars based on \strom colors are affected by 
the presence of molecular bands, such as $CN$, $CH$ and $NH$.
As a matter of fact, two strong cyanogen molecular absorption bands 
are located at $\lambda=4142$ and $\lambda=4215$ \AA, i.e. very close to 
the effective wavelength of the $v$ filter ($\lambda_{eff}=4110$,$\;\; 
\Delta \lambda=190$ \AA). Moreover, the strong $CH$ molecular band 
located in the Fraunhofer's $G-$band ($\lambda=4300$ \AA) might affect 
both the $v$ and the $b$ magnitude. It is noteworthy that the molecular 
$NH$ band  at $\lambda=3360$ \AA, and the two $CN$ bands at $\lambda=3590$ 
and $\lambda=3883$ \AA~ might affect the $u$ ($\lambda_{eff}=3450$,$\;\;
\Delta \lambda=300$ \AA) magnitude (see, e.g. Smith 1987).    
To decrease the contamination by molecular bands in the color index, 
we decided to adopt only colors based on the $y$-band and on the 
NIR bands in our new calibration of MIC relations. The main advantage 
of this approach is that the aforementioned molecular bands only 
affect the \strom $m_1$ index.

We derived theoretical MIC relations based on $m_1$ and the $y$--NIR colors 
based on 2MASS $J,H,K$ filters. Together with the 
classical $m_1$ index, we also computed independent MIC relations for the 
reddening-free parameter $[m] = m_1\, +\, 0.3\, \times (\bmy)$, to overcome 
deceptive uncertainties caused by differential reddening.
To select the $m_1$ and the $[m]$ values along the individual isochrones 
we followed the same approach adopted in CA07.  A multilinear regression fit 
was performed to estimate the coefficients of the MIC relations for the 
$m_1$ and the $[m]$ indices as a function of the three $CI$s, namely 
$y - J$, $y - H$ and $y - K$:

\begin{eqnarray*}
m_1 = \alpha\, + \beta\,\mh + \gamma\, CI +
\delta\, CI^2 +  \epsilon\,  m_1^2 + \zeta\, (CI \times \mh) \\
+  \eta\, (m_1 \times \mh) + \theta\, (CI \times m_1) + \\
\iota\, (CI^2 \times m_1^2) \\
\end{eqnarray*}

where the symbols have their usual meaning. 
To select the form of the analytical relation we followed the forms
adopted for the $m_1$ and the $hk$ metallicity index calibrations in 
CA07 and \citet{io11}, respectively. We performed several tests finding the best solution of the 
multilinear regression fit when adopting the $m_1$ index as an independent variable.  
We then selected the solution with the lowest chi-square of
the multilinear regression fit. As a further check, we estimated the Root Mean Square (RMS)
deviations of the fitted points from the fit and the values range between
0.0005 to 0.0009. Moreover, the multi-correlation parameters attain values close to 1.
The coefficients of the fits, together with their uncertainties, for the 
twelve MIC relations, are listed in Table~2. The RMS values and the 
multi-correlation parameters of the different relations are listed in the 
last two columns of the table.

The above MIC relations are valid in the following color ranges,
0.05 $< m_1 <$ 0.7 mag, 0.2 $< [m] <$ 1.1 mag,
1.4 $< y - J <$ 2.8 mag, 1.8 $< y - H <$ 3.5 mag, and
1.9 $< y - K <$ 3.7 mag, for the scaled-solar models,
while in the following color ranges, 0.05 $< m_1 <$ 0.6 mag, 
0.2 $< [m] <$ 0.9 mag, 1.5 $< y - J <$ 2.7 mag, 2.0 $< y - H <$ 3.5 mag, 
and 2.0 $< y - K <$ 3.7 mag, for the $\alpha$-enhanced models.

\section{Validation of the new metallicity calibration}\label{validation}
\subsection{Field red-giant stars}
In order to validate the new theoretical calibration of the $m_1$ index 
based on visual--NIR colors we estimate the metallicity of field RGs 
for which {\it uvby\/} and NIR photometry and high-resolution spectroscopy are available.
The \strom photometry and the spectroscopy is from \citet[hereafter ATT98]{twa94,twa98}, while
NIR data were retrieved from the 2MASS archive (see CA07 for more details on the selection of
this sample). The sample includes 81 field RGs with a measurement 
in the {\it uvby\/} and the $J,H,K$ bands, a reddening estimate (ATT98)
and high-resolution spectroscopy in the \citet{zinn} metallicity scale. 
For 28 RGs we retrieved \footnote{Data have been retrieved from the VO database using Topcat.} 
the $Calcium$ abundance from \citet{fulbright00}, from which a proxy of the $\alpha$-enhancement 
is estimated, i.e. \afe $\approx [Ca/Fe]$.
We apply the $\alpha$-enhanced calibration since we verified that it is the best suited to
estimate the photometric metallicity of halo field RGs \citep[CA07]{io12}.
The metallicity range covered by current $\alpha$-enhanced MIC relations is
$-2.3 <$ \mh $<$ 0.3 but we select stars with $-2.5 <$ \mh $< $ 0.5 
to account for current uncertainties in spectroscopic abundances and in the GC 
metallicity scale \citep{kra03}. Furthermore, we select the RGs in photometric 
accuracy, i.e. $\sigma_{J,H,K} <$ 0.05 mag, and according to the color range of validity of the 
six MIC relations, ending up with a sample of 61 stars, for 12 of which we have the 
\afe spectroscopic measurement.

To un-redden the $m_1$ index and the colors of the field RGs
we adopt $E(m_1) = - 0.30 \times E(\bmy)$  (CA07), and 
$E(y - J)= 2.23 \times E(B-V)$, 
$E(y - H)= 2.56 \times E(B-V)$,
$E(y - K)= 2.75 \times E(B-V)$, estimated assuming the \citet{card89} 
reddening relation and $R_V = A_V/E(B-V) = 3.1$.

   \begin{figure}
   \includegraphics[width=0.5\textwidth]{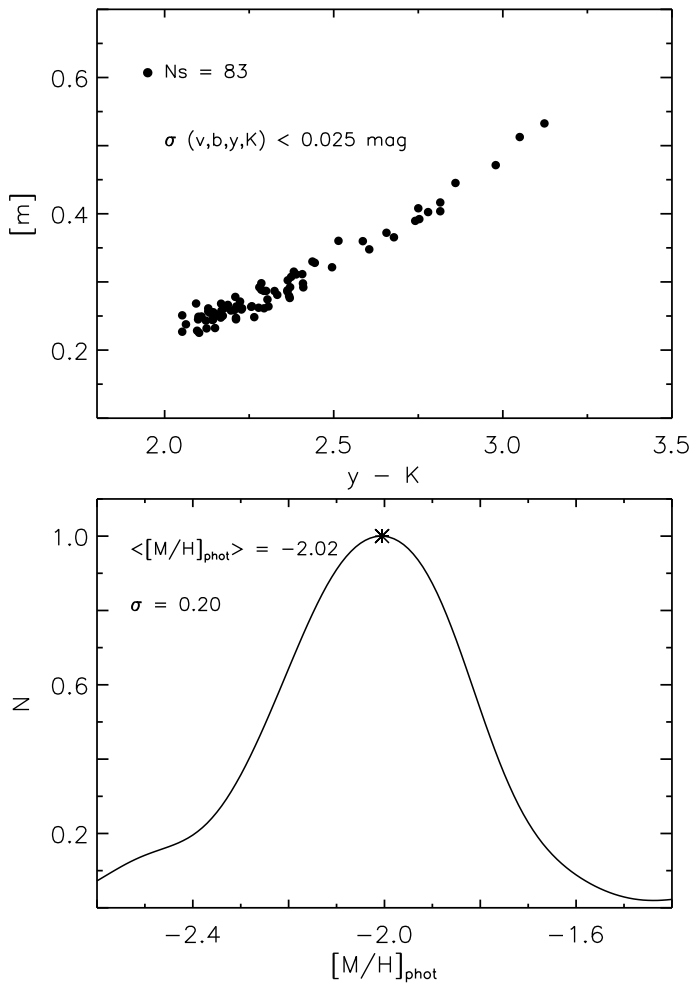}
      \caption{Top: selected RG stars of the GGC M~92 plotted in the 
        $[m], \ y - K$ plane.
	 Bottom: photometric metallicity distribution obtained by applying the 
	 $[m], \ y - K$ MIC relation for the sample 83 RG stars.}
	 \label{fig9}
   \end{figure}
 
The selected stars are plotted in the $[m] ,\ y - J$ plane in panel a) of Fig.~9 (filled dots).  
The star marked with an asterisk is HD~84903, which might be affected by weak chromospheric emission in the
core of the $Ca II K$ line (ATT98). For this star we adopt the spectroscopic
measurement by \citet{thevenin}, estimated by accounting for non-LTE effects \citep{io11}.
The diamond marks a CH-strong star, HD~55496 (ATT98, CA07).
Panel b) of the same figure shows the difference between the photometric and the spectroscopic 
metallicity ($\Delta \mh = (\mh_{\hbox{\footnotesize phot}} - \mh_{\hbox{\footnotesize spec}}))$ 
for the 61 field RG stars as a function of their spectroscopic metal abundances 
(\mh$_{spec}$). We assumed an $\alpha$-enhancement of \afe = 0.4 and estimated
\mh$_{spec}$ = \feh$_{spec}$ + $\log(0.638\times f + 0.362)$, where $\log(f)$ = \afe, for stars with 
no $\alpha$-element abundance measurement. 
The observed dispersion is mainly due to photometric, reddening, and spectroscopic errors. The error bars
in panel b) display the mean error for the spectroscopic abundance measurements 
(see ATT98 and CA07 for more details on how the errors are estimated).
Data plotted in panel b) show that on average there seems to be a shift of photometric metal
abundances towards metal-poor values. 
Photometric metallicities are estimated by adopting the $[m],\ y - J$ MIC relation and
the mean difference between photometric and spectroscopic measurements is $-0.08\pm0.10$ dex, 
with a mean intrinsic dispersion of $\sigma =$ 0.29 dex. The $CH$-strong star (diamond) shows a 
much higher photometric metallicity ($\Delta \mh = \mh_{\hbox{\footnotesize phot}} - \mh_{\hbox{\footnotesize spec}}$ = 0.6) 
as expected (CA07).
The difference between photometric and spectroscopic measurements derived averaging 
the MIC relations based on the reddening free metallicity indices ($[m],\ y-J$, $[m],\ y-H$, $[m],\ y-K$) is -$0.13\pm0.06$ dex. 
Similar results are obtained when applying the other MIC relations, 
and the mean difference derived averaging the $m_1,\ y-J$, $m_1,\ y-H$, $m_1,\ y-K$ relations 
is $-0.16\pm0.05$ dex. In spite of this systematic small shift towards metal-poor values, 
the average intrinsic dispersions are $\sigma =$ 0.25 and $\sigma =$ 0.24 dex, respectively.
Moreover, the shape of the photometric metallicity distribution estimated by adopting the $[m] ,\ y - J$
relation agrees quite well, within uncertainties, 
with the shape of the spectroscopic one (see solid and dashed lines in panel c) ).
A similar agreement between photometric and spectroscopic metallicity distributions is 
obtained when applying the other MIC relations.

A culprit for the systematic shift between photometric and spectroscopic
abundances might be an $\alpha$-enhancement for field stars smaller 
than the assumed $\alpha$ value. 
The evolutionary models adopted to perform our theoretical metallicity calibration 
have been computed assuming \afe = 0.4. This is the typical enhancement 
found in cluster stars using high-resolution spectra \citep{kra94, gratton}. Unluckily, 
we do not have an $\alpha$-enhancement abundance measurement for our field star sample but
only an estimate of the $\alpha$-enhancement proxy, i.e. \afe $\approx [Ca/Fe]$, 
for 12 out of the 61 RGs, with a median value of \afe = 0.35, and a dispersion of 0.08 dex.
It is worth mentioning that we were finding a similar shift, $\approx -0.1$ dex, 
when adopting the visual--NIR $\alpha$-enhanced MIC relations to estimate photometric 
metallicities of field dwarfs in \citet{io12}. 
When assuming an $\alpha$-enhancement of 0.2 for the stars without the \afe
spectroscopic estimate, we find a very good agreement between spectroscopic and 
photometric abundance estimates. We obtain a mean difference of 
$0.04\pm0.05$ dex, with a mean intrinsic dispersion of $\sigma =$ 0.30 dex 
($[m],\ y-J$, $[m],\ y-H$, $[m],\ y-K$) and  $0.00\pm0.05$ dex, 
with $\sigma =$ 0.30 dex ( $m_1,\ y-J$, $m_1,\ y-H$, $m_1,\ y-K$). 
On the other hand, when adopting the scaled-solar calibration to estimate the abundance of the
same sample we obtain photometric metallicities systematically
more metal-poor than the spectroscopic abundances, with a mean difference of $-0.19\pm$0.04 
and a mean dispersion of $\sigma =$ 0.26 dex for the $[m],\ CI$ relations, while a mean difference of 
$-0.24\pm$0.02 and a mean dispersion of $\sigma =$ 0.21 dex for the $m_1,\ CI$ relations.

  \begin{figure}
   \includegraphics[width=0.5\textwidth]{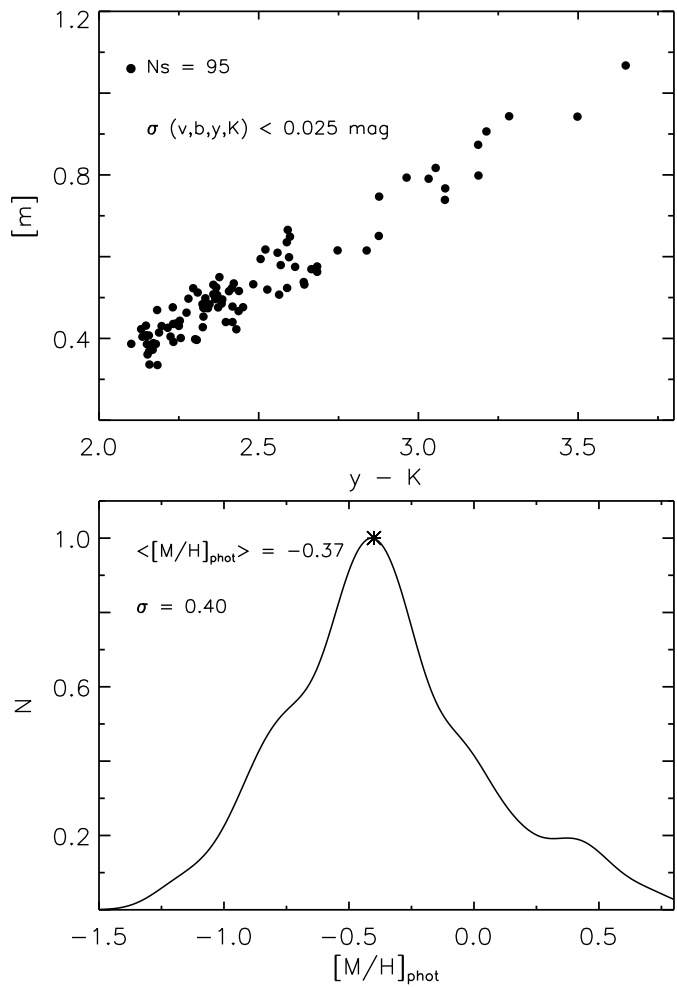}
      \caption{Top: selected RGs of the bulge cluster NGC~6624 plotted in the 
        $[m], \ y - K$ plane.
	 Bottom: photometric metallicity distribution obtained by applying the 
	 $[m], \ y - K$ MIC relation for the sample of 95 RGs.}
	 \label{fig10}
   \end{figure}

     \begin{figure}
   \includegraphics[width=0.5\textwidth]{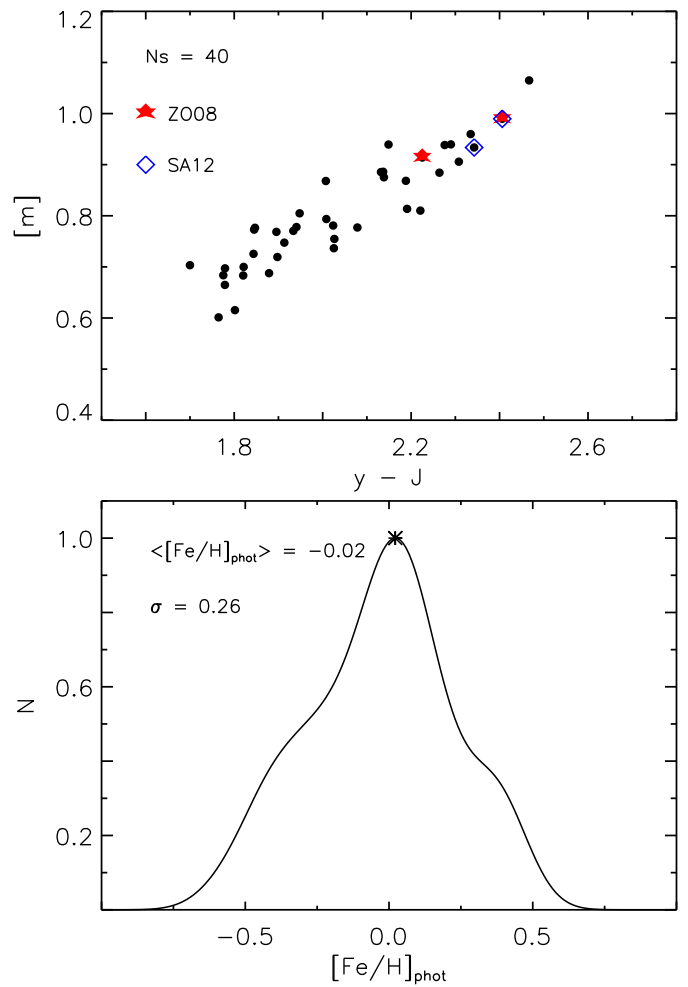}
      \caption{Top: selected RGs of the bulge cluster NGC~6528 plotted in the 
        $[m], \ y - J$ plane.
	 Bottom: photometric metallicity distribution obtained by applying the 
	 $[m], \ y - J$ MIC relation for the sample of 40 RGs.
	 Stars with spectroscopic measurements by ZO08 and SA12 are marked
	 with filled red stars and open blue diamonds, respectively.
	 }
	 \label{fig11}
   \end{figure}
   
   \begin{figure}
   \includegraphics[width=0.5\textwidth]{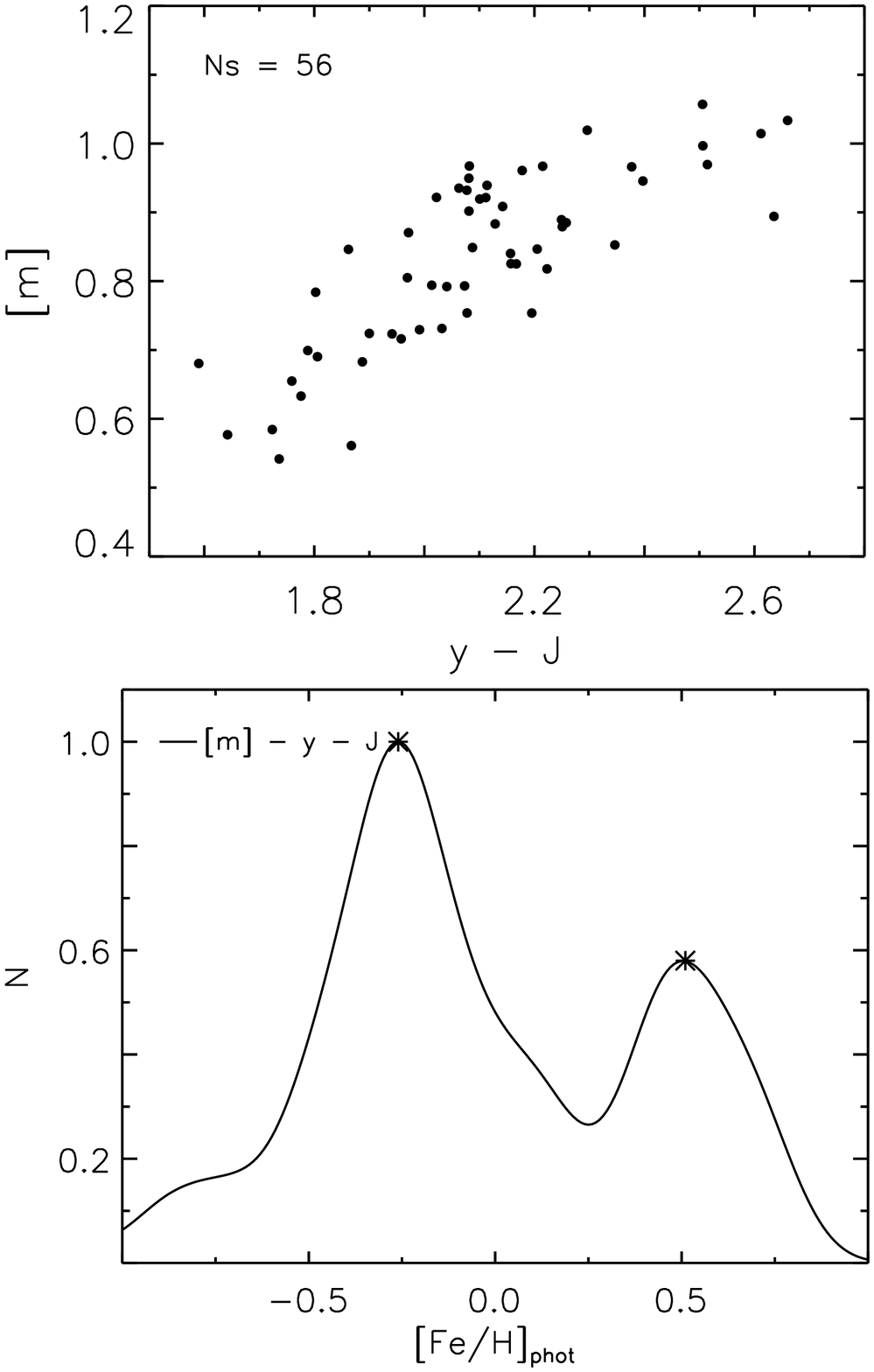}
      \caption{Top: selected RGs of the field surrounding NGC~6528 in the Baade's window 
        plotted in the $[m], \ y - J$ plane.
	 Bottom: photometric metallicity distribution obtained by applying the 
	 $[m], \ y - J$ MIC relation to the sample of 56 Baade's window RGs.
	 Stars with spectroscopic measurements by GO11 are marked
	 with filled red stars.
	 }
	 \label{fig12}
   \end{figure}


\subsection{Cluster red-giant stars}
To further validate the new theoretical visual--NIR metallicity calibration, we also
adopt the photometry of a metal-poor halo GGC, M~92 (NGC~6341, \feh = $-2.35\pm$0.05, \citealt{carretta09}) 
and of a metal-rich bulge cluster, NGC~6624 (\feh = $-0.69\pm$0.02, \citealt{valenti11}).

The use of cluster data brings forward three indisputable advantages: 

{\em a)} The evolutionary status (age, effective temperature, surface 
gravity, stellar mass) of cluster RGs is well established.

{\em b)} The abundance of iron and $\alpha$-elements is known with 
high-precision.

{\em c)} The selected clusters cover a broad range in iron abundance. 

Data for NGC~6624 were collected with EFOSC2 during July 12 and 13 2010. 
We have $u$-, 8$v$-, 6$b$- and 4$y$-band images, 
with exposure times ranging between $40$ to $380$s and seeing between 1\farcs0 to 1\farcs5. 
The total FoV is 4.6\min$\times$4.6\min and includes the cluster center. We followed the same
approach adopted for NGC~6528 in reducing and calibrating the data, ending up
with a catalog of  $6,217$ stars. These data were matched with NIR VIRCAM photometry,
obtaining an optical--NIR catalog of $4,252$ stars. 

The \strom catalog of M~92 was obtained with images collected with the 2.56m Nordic Optical Telescope (NOT) 
on La Palma (for more details see \citealt[CA07]{gru00}). We cross-correlated the \strom
photometry of M~92 with the NIR photometry of \citet{valenti04} for the RGs in this cluster, 
ending up with a sample of 340 stars with a measurement in the $u,v,b,y$ and $J,K$ bands.
We then adopted the $b - J,\ y - K$ color--color plane to disentangle the M~92 RGs from field 
stars as described in CA07, obtaining a clean cluster sample of 223 stars.
We applied the same procedure to disentangle NGC~6624 RGs from field stars,
ending up with a sample of 147 RGs for this cluster.

The top panel of Fig.~10 shows M~92 candidate member RGs selected in photometric accuracy, 
i.e. $\sigma_{v,b,y,K} <$ 0.025 mag, and plotted in the $[m],\ y - K$ plane;
We adopted  $E(B-V)$= 0.02 mag for M~92 \citep{dicecco} and the relations derived in \S 5 to
un-redden the cluster photometry.
The bottom panel shows the metallicity distribution obtained for the 86 RGs by adopting the $[m],\ y - K$ $\alpha$-enhanced MIC relation.
The distribution was smoothed by applying a Gaussian kernel having 
a standard deviation equal to the photometric error in the $m_1$ index, following the prescriptions of \citet{io09, io12}.
The distribution has a quite symmetric shape, and most of the spread is due to photometric errors. 
The residual spread is due to the effect of $CN/CH/NH$ molecular bands on the $m_1$ index;
M~92 shows the typical variations in $[C/Fe]$ and $[N/Fe]$  \citep{carbon82,langer,bellman}, together with 
the usual anti-correlations of most GGCs \citep{pila, sneden, kra94}.
We also found a similar effect when applying the visual--NIR metallicity calibration to M~92 MS stars \citep{io12}.

By fitting it with a Gaussian we obtain a main peak at \mh = $-2.02$ (\feh = $-2.37$), 
with a dispersion of $\sigma =$ 0.20 dex. The metallicity distributions obtained by applying the other 
MIC relations agree with each other, with averaged mean peaks of -2.05$\pm$0.04 (\feh = $-2.40$) 
and a mean intrinsic dispersion of $\sigma =$ 0.31 dex ($[m],\ y - J$, 
$[m],\ y - K$ relations), and of -2.07$\pm$0.01 (\feh = $-2.42$), 
with $\sigma =$  0.25 dex ($m_1,\ y - J$, $m_1,\ y - K$ relations).
These estimates are in good agreement, within uncertainties, with spectroscopic 
estimates from the literature (\feh = $-2.24\pm$0.10, \citealt{zinn}, and \feh = $-2.35\pm$0.05, \citealt{carretta09}).
On the other hand, when applying the scaled-solar MIC relations to estimate the iron abundance of 
M~92, we obtain a mean peak of (\feh = $-2.17\pm$0.01, and a mean intrinsic dispersion 
of $\sigma =$ 0.24 dex ($[m],\ y - J$,  $[m],\ y - K$ relations), and of $-2.15\pm$0.02,
with $\sigma =$  0.40 dex ($m_1,\ y - J$, $m_1,\ y - K$ relations). These values are on average
slightly ($\sim$ 0.1 dex) more metal-rich than the spectroscopic values.
 
Fig.~11 shows the same test performed for NGC~6624. 
We adopted a reddening of $E(B-V)$= 0.28 mag for this cluster \citep{valenti04}.
The candidate RGs clearly split in two sequences in the $[m],\ y - K$ plane (see top panel).
The split is due to the effect of the molecular bands, such as $CN$, $CH$ and $NH$, on the $m_1$ index, 
that is more pronounced in metal-rich clusters, as shown in CA07, and constrained
by ACS-HST photometry in the case of NGC~1851 and 47 Tuc \citep{milone08, milone12}.
The presence of  this dichotomy in NGC~6624 has not yet been confirmed by spectroscopic studies.

The smoothed metallicity distribution obtained by applying the $[m],\ y - K$ $\alpha$-enhanced MIC relation is 
shown in the bottom panel of Fig.~11. The shape of the distribution is asymmetric, with a tail towards 
the metal-rich regime up to \mh $\approx$ 0.7, and a secondary peak at \mh $\approx$ 0.4. The metal-rich
tail could be due to the residual contamination by field disk stars. 
Fitting the distribution with a single Gaussian we obtain a main peak at \mh = $-0.37$ (\feh = $-0.72$), with a dispersion of
$\sigma =$ 0.40 dex. Most of the dispersion is due to photometric errors, but part of it is due to the presence 
of uncorrected differential reddening and the rest is intrinsic, i.e. given to the presence of $CN$-,$CH$-strong stars.
The metallicity distributions obtained by applying the other MIC relations agree with each other, 
with averaged mean peaks of $-0.39\pm$0.09 (\feh = $-0.74$) and a mean intrinsic dispersion of $\sigma =$ 0.42 dex 
($[m],\ y - J$, $[m],\ y - H$, and $[m],\ y - K$ relations), and of $-0.40\pm$0.04 (\feh = $-0.75$), 
with $\sigma =$  0.42 dex ($m_1,\ y - J$, $m_1,\ y - H$, and $m_1,\ y - K$ relations).
These estimates are in good agreement, within uncertainties, with high-resolution spectroscopic 
estimates from  \citet{valenti11}, \feh = $-0.69\pm$0.02, and with the CaT measurements
of  \citet{heasley}, \feh = $-0.63\pm$0.09, and of \citet{mauro14}, \feh = $-0.67\pm$0.10.
On the other hand, by applying the scaled-solar metallicity relations to estimate the cluster iron abundance we obtain 
an average mean peak of \feh = $-0.97\pm$0.03 and a mean intrinsic dispersion of $\sigma =$ 0.28 dex 
($[m],\ y - J$, $[m],\ y - H$, and $[m],\ y - K$ relations), and of $-0.99\pm$0.06, 
with $\sigma =$  0.34 dex ($m_1,\ y - J$, $m_1,\ y - H$, and $m_1,\ y - K$ relations). 
These iron abundances are almost 0.3 dex more metal-poor than the spectroscopic values.


\section{The metallicity of the bulge cluster NGC~6528 and its surrounding field.}

We adopt the new visual--NIR calibration to estimate the metal abundance of 
the bulge cluster NGC~6528 and to study the metallicity distribution of RGs 
in the surrounding field in the Baade's window.

We select 97 RG stars from the clean cluster sample shown in Fig.~2.
Stars are further selected in photometric accuracy, i.e. $\sigma_{v,b,y,J,H,K}  \le$ 0.025 mag, 
separation index, $sep_{v,b,y} \ge 5$, and for the color ranges of validity of the calibration, 
obtaining a final sample of 40 NGC~6528 RGs. We adopted $E(B-V)$ = 0.54 mag 
to un-redden the data (see \S 4). The top panel of 
Fig.~12 shows the selected stars plotted in the $[m],\ y - J$ plane. The two filled red stars mark RGs with 
high-resolution spectroscopic measurement from ZO08 and the empty blue diamonds mark two RGs with 
$CaT$ spectroscopic abundances from SA12. One star, I~42, belongs to
both samples. Four stars from the sample of SA12 were rejected since they did not pass the selection in 
photometric accuracy or they were outside the color range of validity of the calibration.
The dispersion of the sequence in the $[m],\ y - J$ plane is mostly due to photometric errors, 
and possibly to some residual uncorrected differential reddening. It is noteworthy that RGs do not split
 in two parallel sequences in this color--color plane, as in the case of NGC~6624.  
Unfortunately, we do not have enough statistics to ascertain if the absence of the split is intrinsic or given 
to the effect of the paucity of the sample. We also do not have enough spectroscopic data to 
constrain the effect of $CN/CN/NH$ molecular bands on the broadening of the RG sequence
in the $[m],\ y - J$ plane.

The smoothed metallicity distribution obtained by applying the scaled-solar $[m],\ y - J$ MIC relation is shown
in the bottom panel of Fig.~12. The shape of the distribution is quite similar to the shape of the 
metallicity distribution of NGC~6624, while the dispersion is much lower, 
$\sigma$ = 0.19 dex compared to $\sigma$ = 0.40 dex.
Part of the difference could be due to the fact that NGC~6528 \stromc--NIR catalog was corrected for 
differential reddening and cleaned from the contamination of field disk stars by using proper-motion measurements
estimated with HST photometry, while NGC~6624 catalog was cleaned by field contamination only by adopting the
\stromc--NIR color--color planes, and it was not corrected for differential reddening. 
The metal-rich tail present in the NGC~6624 metallicity distribution is indeed absent 
in the NGC~6528 distribution, but the distribution does not look to be mono-parametric.
Fitting NGC~6528 MDF with a single Gaussian we obtain a 
main peak at \feh = $-0.02$, with a dispersion of $\sigma =$ 0.19 dex, by adopting the scaled-solar $[m],\ y - J$ MIC relation.
The metallicity distributions obtained by applying the other scaled-solar MIC relations agree with each other, 
with average mean peaks of $-0.04\pm$0.02 and a mean intrinsic dispersion of $\sigma =$ 0.27 dex 
($[m],\ y - J$ and $[m],\ y - K$ relations), and of $-0.11\pm$0.01, with $\sigma =$  0.27 dex 
($m_1,\ y - J$ and $m_1,\ y - K$ relations).
The photometric metallicity estimates for the four RGs with the corresponding spectroscopic measurements from ZO08 
and SA12 are listed in Table~3, together with the spectroscopic measurement errors. 
For star I~42 we estimated $\alpha$-enhancement as \afe = $[Ca+ Mg + Si/Fe]$. 
The reduced equivalent widths -- $W'$ -- from SA12 were converted to \feh measurements 
by adopting the relation presented in the paper (the interested reader is referred to SA12 for more details), 
and then converted to the \citet{zinn} metallicity scale by adopting the relation provided by \citet{carretta09}.
The agreement between photometric and spectroscopic estimates for the four RGs 
is very good within the uncertainties (see Table~3).

We adopted also the $\alpha$-enhanced MIC relations to estimate the metallicities of NGC~6528 RGs but we 
obtained metal abundances of more than 0.5 dex too metal-rich, i.e. 
$\Delta \mh = \mh_{\hbox{\footnotesize phot}} - \mh_{\hbox{\footnotesize spec}} \approx$ 0.5 dex.


We then adopted our clean sample of bulge stars to study the metallicity distribution of the
Baade's window. We selected RG stars from the sample shown in Fig.~4, ending up with 62 candidate bulge RGs.
The stars are then further selected in photometric accuracy, i.e. $\sigma_{v,b,y,J,H,K}  \le$ 0.03 mag and 
separation index, $sep_{v,b,y} \ge$ 0, and for the color ranges of validity of the calibration, 
obtaining a final sample of 56 RGs. The sample photometry has been un-reddened by using the same
reddening adopted for NGC~6528. The top panel of Fig.~12  shows the selected stars plotted 
in the $[m],\ y - J$ plane. 
The bottom panel of Fig.~12 displays the smoothed photometric metallicity distribution
obtained by applying the scaled-solar $[m], \ y - J$ MIC relation. The distribution is clearly bimodal, with a main 
peak at \feh $\approx -0.25$ and a secondary one at \feh $\approx$ 0.5. Similar distributions 
are obtained when adopting the other visual--NIR MIC relations, with average peaks at 
\feh $\approx -0.2$ and \feh $\approx$ 0.55 ($[m],\ y - J$, $[m],\ y - H$ and $m_1,\ y - K$ relations),
and \feh $\approx -0.25$ and \feh $\approx$ 0.4 ($m_1,\ y - J$, $m_1,\ y - H$ , and $m_1,\ y - K$ relations).
By applying the $\alpha$-enhanced MIC relations we obtain a metallicity distribution shifted 
by almost 1 dex towards the metal-rich regime. 


\begin{table*}
\caption{Log of the \strom images collected with EFOSC2 on the NTT for the bulge 
cluster NGC~6528 and the halo cluster NGC~6752 adopted in this investigation (program ID: 085.D-0374, PI: A. Calamida).}      
\label{table:1}      
\begin{tabular}{l c c c c c }        
\hline\hline                 
Name & Exposure time & Filter & RA & DEC & Seeing \\    
     &  (s)          &        & (hh:mm:ss.s) & (dd:mm:ss.s) & (arcsec) \\  
\hline                        
\hline
NGC~6528\\
\hline
July 11, 2010\\
\hline
EFOSC.2010-07-12T03:49:59.514.fits  & 60   & y & 18:04:42.6 & -30:04:56.9 & 0.8 \\
EFOSC.2010-07-12T03:51:42.194.fits  & 200  & b & 18:04:42.6 & -30:04:56.9 & 0.8 \\
EFOSC.2010-07-12T03:55:45.076.fits  & 1000 & v & 18:04:42.6 & -30:04:56.9 & 0.8 \\
EFOSC.2010-07-12T04:13:08.053.fits  & 2000 & u & 18:04:42.6 & -30:04:56.9 & 1.1 \\
\hline
July 13, 2010 \\
\hline
EFOSC.2010-07-14T04:03:08.983.fits & 60   & y & 18:04:38.5 & -30:05:57.0 & 1.1 \\
EFOSC.2010-07-14T04:04:47.581.fits & 60   & y & 18:04:37.4 & -30:06:11.9 & 1.1 \\
EFOSC.2010-07-14T04:06:44.945.fits & 300  & b & 18:04:38.5 & -30:05:57.0 & 1.1 \\
EFOSC.2010-07-14T04:12:27.376.fits & 300  & b & 18:04:37.4 & -30:06:11.9 & 1.2 \\
EFOSC.2010-07-14T04:18:18.498.fits & 2500 & v & 18:04:38.5 & -30:05:57.0 & 1.5 \\
EFOSC.2010-07-14T05:01:01.323.fits & 2100 & u & 18:04:38.5 & -30:05:57.0 & 1.7 \\
\hline                                   
\hline
NGC~6752\\
\hline
July 11, 2010\\
\hline
EFOSC.2010-07-12T03:14:57.815.fits & 3 & y & 19:10:53.6 & -59:59:21.4 & 1.2 \\
EFOSC.2010-07-12T03:15:41.972.fits & 3 & y & 19:10:51.7 & -59:58:55.4 & 1.2 \\
EFOSC.2010-07-12T03:16:27.729.fits & 3 & y & 19:10:49.6 & -59:59:21.3 & 1.2 \\
EFOSC.2010-07-12T03:17:35.375.fits & 6 & b & 19:10:53.6 & -59:59:21.4 & 1.2 \\
EFOSC.2010-07-12T03:18:23.873.fits & 6 & b & 19:10:51.7 & -59:58:55.4 & 1.2 \\
EFOSC.2010-07-12T03:19:11.922.fits & 6 & b & 19:10:49.6 & -59:59:21.3 & 1.2 \\
EFOSC.2010-07-12T03:20:21.398.fits & 30 & v & 19:10:53.6 & -59:59:21.4 & 1.2 \\
EFOSC.2010-07-12T03:21:34.085.fits & 30 & v & 19:10:51.7 & -59:58:55.4 & 1.2 \\
EFOSC.2010-07-12T03:22:44.113.fits & 30 & v & 19:10:49.6 & -59:59:21.3 & 1.2 \\
EFOSC.2010-07-12T03:24:17.518.fits & 60 & u & 19:10:53.6 & -59:59:21.4 & 1.2 \\
EFOSC.2010-07-12T03:26:00.197.fits & 60 & u & 19:10:51.7 & -59:58:55.4 & 1.2 \\
EFOSC.2010-07-12T03:27:42.216.fits & 60 & u & 19:10:49.6 & -59:59:21.3 & 1.2 \\
\hline
July 13, 2010\\
\hline
EFOSC.2010-07-14T03:31:07.342.fits & 10 & y & 19:10:53.6 & -59:59:21.4 & 1.2 \\
EFOSC.2010-07-14T03:31:59.032.fits & 10 & y & 19:10:51.7 & -59:58:55.4 & 1.2 \\
EFOSC.2010-07-14T03:32:49.501.fits & 10 & y & 19:10:49.6 & -59:59:21.3 & 1.2 \\
EFOSC.2010-07-14T03:34:01.889.fits & 25 & b & 19:10:53.6 & -59:59:21.4 & 1.2 \\
EFOSC.2010-07-14T03:35:06.773.fits & 25 & b & 19:10:51.7 & -59:58:55.4 & 1.2 \\
EFOSC.2010-07-14T03:36:11.228.fits & 25 & b & 19:10:49.6 & -59:59:21.3 & 1.2 \\
EFOSC.2010-07-14T03:37:37.140.fits & 60 & v & 19:10:53.6 & -59:59:21.4 & 1.2 \\
EFOSC.2010-07-14T03:39:18.139.fits & 60 & v & 19:10:51.7 & -59:58:55.4 & 1.2 \\
EFOSC.2010-07-14T03:40:58.507.fits & 60 & v & 19:10:49.6 & -59:59:21.3 & 1.2 \\
EFOSC.2010-07-14T03:42:58.503.fits & 80 & u & 19:10:53.6 & -59:59:21.4 & 1.2 \\
EFOSC.2010-07-14T03:45:01.599.fits & 80 & u & 19:10:51.7 & -59:58:55.4 & 1.2 \\
EFOSC.2010-07-14T03:47:01.975.fits & 80 & u & 19:10:49.6 & -59:59:21.3 & 1.2 \\
\hline
\hline
\end{tabular}
\tablefoot{This table is available in its entirety in a machine-readable form in the online journal.}
\end{table*}

\begin{table*}
\caption{Multilinear regression coefficients for the \strom  
metallicity index: $m_1 = \alpha\, + \beta\,\feh + \gamma\, CI +
\delta\, CI^2 + \epsilon\,m_1^2 + \zeta\, (CI \times \feh) + \eta\, (m_1 \times \feh) 
+ \theta\, (CI \times m_1) + \iota\, (CI^2 \times m_1^2)$}             
\label{table:2}      
\begin{tabular}{l c c c c c c c c c c c}        
\hline\hline                 
Relation & $\alpha$ & $\beta$ & $\gamma$ & $\delta$ & $\epsilon$ & $\zeta$ & $\eta$ & $\theta$ & $\iota$ & Multicorr & RMS \\    
\hline                        
Scaled-solar\\
\hline
$m_1,  y - J$ & -0.107 & 0.133 & 0.365 & -0.196 & -0.715 & -0.090 & 0.130 & 0.860 & -0.012 & 1.000 & 0.0007 \\    
Error         &  0.002 & 0.007 & 0.004 &  0.043 &  0.002 &  0.006 & 0.018 & 0.001 &  0.006 & $(\ldots)$ & $(\ldots)$ \\    
$[m],  y - J$ & -0.138 & 0.132 & 0.500 & -0.319 & -0.918 & -0.111 & 0.161 & 1.132 & -0.004 & 1.000 & 0.0008 \\   
Error         &  0.002 & 0.007 & 0.005 &  0.041 &  0.002 &  0.006 & 0.026 & 0.001 &  0.005 & $(\ldots)$ & $(\ldots)$ \\
$m_1,  y - H$ & -0.060 & 0.132 & 0.237 & -0.104 & -0.662 & -0.068 & 0.133 & 0.649 & -0.012 & 1.000 & 0.0007 \\   
Error         & 0.0014 & 0.004 & 0.002 &  0.042 &  0.001 &  0.006 & 0.013 & 0.001 &  0.004 & $(\ldots)$ & $(\ldots)$ \\
$[m],  y - H$ & -0.103 & 0.137 & 0.361 & -0.174 & -0.528 & -0.080 & 0.093 & 0.721 & -0.006 & 1.000 & 0.0009 \\    
Error         & 0.0014 & 0.007 & 0.003 &  0.040 &  0.001 &  0.008 & 0.018 & 0.001 &  0.006 & $(\ldots)$& $(\ldots)$ \\    
$m_1,  y - K$ & -0.068 & 0.137 & 0.239 & -0.103 & -0.719 & -0.070 & 0.137 & 0.650 & -0.009 & 1.000 & 0.0008 \\   
Error         & 0.0015 & 0.004 & 0.002 &  0.043 &  0.001 &  0.007 & 0.013 & 0.001 &  0.004 & $(\ldots)$ & $(\ldots)$ \\
$[m],  y - K$ & -0.116 & 0.149 & 0.363 & -0.155 & -0.454 & -0.077 & 0.084 & 0.637 & -0.003 & 1.000 & 0.0008 \\    
Error         & 0.0015 & 0.007 & 0.003 &  0.045 &  0.001 &  0.008 & 0.021 & 0.001 &  0.007 & $(\ldots)$& $(\ldots)$ \\    
\hline
\hline
$\alpha$-enhanced\\
\hline
$m_1,  y - J$ & -0.123 & 0.091 & 0.310 & -0.137 & -0.516 & -0.053 & 0.075 & 0.687 & -0.019 & 1.000 & 0.0005 \\
Error         &  0.001 & 0.005 & 0.003 &  0.050 &  0.001 &  0.005 & 0.016 & 0.002 &  0.004 & $(\ldots)$ & $(\ldots)$ \\
$[m],  y - J$ & -0.148 & 0.091 & 0.438 & -0.232 & -0.628 & -0.065 & 0.084 & 0.869 & -0.009 & 0.998 & 0.0006 \\
Error         &  0.001 & 0.009 & 0.008 &  0.077 &  0.002 &  0.008 & 0.041 & 0.002 &  0.008 & $(\ldots)$ & $(\ldots)$ \\
$m_1,  y - H$ & -0.094 & 0.099 & 0.220 & -0.079 & -0.467 & -0.045 & 0.077 & 0.532 & -0.014 & 1.000 & 0.0005 \\
Error         &  0.008 & 0.003 & 0.002 &  0.054 &  0.001 &  0.006 & 0.014 & 0.001 &  0.004 & $(\ldots)$ & $(\ldots)$ \\
$[m],  y - H$ & -0.131 & 0.093 & 0.331 & -0.145 & -0.550 & -0.054 & 0.071 & 0.684 & -0.008 & 1.000 & 0.0005 \\
Error         &  0.001 & 0.004 & 0.003 &  0.039 &  0.001 &  0.005 & 0.016 & 0.005 &  0.004 & $(\ldots)$ & $(\ldots)$ \\
$m_1,  y - K$ & -0.101 & 0.096 & 0.217 & -0.076 & -0.595 & -0.044 & 0.090 & 0.543 & -0.011 & 0.999 & 0.0005 \\
Error         &  0.001 & 0.003 & 0.001 &  0.050 &  0.001 &  0.005 & 0.012 & 0.001 &  0.003 & $(\ldots)$ & $(\ldots)$ \\
$[m],  y - K$ & -0.138 & 0.095 & 0.322 & -0.127 & -0.421 & -0.048 & 0.058 & 0.596 & -0.008 & 0.999 & 0.0005 \\
Error         &  0.001 & 0.004 & 0.002 &  0.042 &  0.001 &  0.005 & 0.017 & 0.001 &  0.004 & $(\ldots)$ & $(\ldots)$ \\
\hline                                   
\end{tabular}
\end{table*}

\begin{table*}
\caption{Photometric and spectroscopic metallicities for four RG stars in the bulge cluster NGC~6528.}      
\label{table:3}      
\begin{tabular}{llcccccc}        
\hline\hline                 
Name   &  $[Fe/H]_{m_1,\ yJ}$ & $[Fe/H]_{[m],\ yJ}$ & $[Fe/H]_{m_1,\ yK}$ &$[Fe/H]_{[m],\ yK}$ &$[Fe/H]_{s}$ & $err([Fe/H]_{s})$ & $[\alpha/Fe]$ \\    
\hline                        
357459 & -0.09 &  0.04 & -0.24      & -0.09      & -0.01\tablefootmark{a}  & 0.29   & $(\ldots)$ \\
I42 & -0.03 & -0.04 & 0.26 & -0.10  & -0.14\tablefootmark{a}   & 0.075  &  0.09\tablefootmark{a} \\
R1-42 & -0.03 & -0.04 & 0.26      & -0.10      & -0.05\tablefootmark{b} & 0.15   &  0.09\tablefootmark{a} \\
R2-41 &  0.05 & -0.19 &  0.18      &  0.10      & -0.15\tablefootmark{b} & 0.15   & $(\ldots)$ \\
\hline                                   
\end{tabular}
\tablefoot{\hspace*{0.5mm}$^a$ High-resolution spectroscopic measurement from ZO08.
\hspace*{0.5mm}$^b$  $CaT$ spectroscopic measurement from SA12.
}
\end{table*}

\section{Summary and conclusions}

We presented \stromc--NIR photometry of the the bulge globular cluster
NGC~6528 and its surrounding field in the Baade's window. 
The main findings concerning \stromc--NIR photometry of cluster and field stars are the following: 

{\em a\/}) The isochrone fit of the proper-motion-cleaned and differential reddening 
corrected \stromc--NIR CMDs for NGC~6528 suggests an age of 11$\pm$1 Gyr, 
by adopting a scaled-solar isochrone with solar abundance, i.e. $Z = 0.0198$, 
$Y = 0.273$, or a $\alpha$-enhanced isochrone with the same iron content, i.e. $Z = 0.04$, $Y = 0.303$.
The same scaled-solar and $\alpha$-enhanced isochrones agree 
quite well, within the uncertainties, with the NIR CMD of NGC~6528.
This result is in good agreement with literature age estimates, i.e. 
$t =$ 11$\pm$2 Gyr \citep{feltz02}, 12.6 Gyr \citep{momany03}, and 11$\pm$1 Gyr (LA14).

{\em b)} We adopted the same theoretical framework used for NGC~6528 to fit 
\stromc--NIR CMDs of proper-motion selected and differential reddening 
corrected bulge stars. We find that a fraction of observed stars 
along the MS and the RGB are systematically redder and/or fainter 
than predicted by adopted cluster isochrones. This result would suggest 
the possible occurrence of large samples of super metal--rich stars, \feh $>$ 0.1,
in the Baade's window.
However, these objects could also be explained with an increase either 
in differential reddening or in depth or both.

{\em c)} We find that central helium burning stars display a well defined 
double peak distribution. The fainter peak is located at K$\sim$13.25$\pm$0.02
and appears to be associated with the RHB stars, since the peak in the $K$-band 
luminosity function of cluster RGs in located at K$\sim$13.16$\pm$0.02 mag.   
The brighter (K$\sim$12.86$\pm$0.02) peak seems to be associated with 
intermediate-age RC stars. The two peaks also show a clear separation in color 
(RHB, J-K$\sim$1.0; RC, J-K$\sim$0.9). More quantitative constraints 
concerning the age and the metallicity distribution of the above evolutionary 
features do require detailed synthetic CMDs and luminosity functions 
(Calamida et al., in preparation).  

Furthermore, we also provided a new theoretical metallicity calibration based on the $m_1$
index and on visual--NIR colors to estimate the global metal abundance of cluster and field RG stars.
We adopted scaled-solar and $\alpha$-enhanced evolutionary models.
This is the first time that visual--NIR colors are adopted to 
estimate photometric metallicities of RG stars. 

We validated the new theoretical metallicity calibration by adopting a sample of field RGs 
with \strom and NIR photometry and high-resolution spectroscopy available. 
The sample includes 61 RGs selected from the study by ATT94 and ATT98.
The mean difference between photometric and spectroscopic abundance
is $-0.16\pm0.05$ dex, with a mean intrinsic dispersion of $\sigma =$ 0.24 dex 
($m_1,\ y-J$, and, $m_1,\ y-K$ relations),
and $0.13\pm0.06$ dex, with $\sigma =$0.25 dex ($[m],\ y-J$ and  $[m],\ y-K$ relations).

The quoted independent comparisons indicate that the new theoretical 
MIC relations provide accurate metal abundances for field RG stars 
with a dispersion smaller than 0.3 dex.

We also tested the calibration by adopting RG stars of two GGCs covering a broad 
range in metal abundance, i.e. M~92 (\feh =  $-2.31$) and NGC~6624 (\feh = $-0.69$), 
for which both \strom and NIR photometry were available. 

We find that the metallicity distributions of M~92 based on the $\alpha$-enhanced
visual--NIR MIC relations are in very good agreement with the spectroscopic estimates 
available in literature, i.e. \feh = $-2.24\pm$0.10 \citep{zinn}, and \feh = $-2.35\pm$0.05
\citep{carretta09}. We find, indeed, a global metallicity of  \mh = $-2.05\pm$0.04 (\feh = $-2.40$), 
with a mean intrinsic dispersion of $\sigma =$ 0.31 dex, by averaging the values obtained by 
adopting the $[m],\ y - J$ and the $[m],\ y - K$ relations, and of $-2.07\pm$0.01 (\feh = $-2.42$), 
with $\sigma =$  0.25 dex, by using the $m_1,\ y - J$ and $m_1,\ y - K$ relations.

By applying the $\alpha$-enhanced visual--NIR MIC relations to the bulge cluster NGC~6624 
we find averaged mean peaks of $-0.39\pm$0.09 (\feh = $-0.74$) and a mean intrinsic dispersion of $\sigma =$ 0.42 dex 
($[m],\ y - J$, $[m],\ y - H$, and $[m],\ y - K$ relations), and of $-0.40\pm$0.04 (\feh = $-0.75$), 
with $\sigma =$  0.42 dex ($m_1,\ y - J$, $m_1,\ y - H$, and $m_1,\ y - K$ relations).
These estimates are in good agreement, within uncertainties, with spectroscopic 
estimates from the literature (\feh = $-0.63\pm$0.09, \citealt{heasley}, and \feh = $-0.69\pm$0.02, \citealt{valenti11}).

We also apply the new calibration to study the metallicity distribution of the bulge cluster
NGC~6528 and its surrounding field. 
The smoothed metallicity distribution obtained by applying the scaled-solar $[m],\ y - J$ MIC 
relation to the proper-motion selected and differential reddening corrected cluster RGs shows a
main peak at \feh = $-0.02$, with a dispersion of $\sigma =$ 0.19 dex.
The metallicity distributions obtained by applying the other scaled-solar MIC relations agree with each other, 
with average mean peaks of $-0.04\pm$0.02 and a mean intrinsic dispersion of $\sigma =$ 0.27 dex 
($[m],\ y - J$ and $[m],\ y - K$ relations), and of $-0.11\pm$0.01, with $\sigma =$  0.27 dex 
($m_1,\ y - J$ and $m_1,\ y - K$ relations).
The agreement between photometric and spectroscopic estimates for the four RGs in common with the
spectroscopic samples of ZO04 and SA12 is, within the uncertainties, very good.
By adopting the $\alpha$-enhanced MIC relations to estimate the metallicities of NGC~6528 RGs 
we obtain metal abundances of more than 0.5 dex too metal-rich, i.e. 
$\Delta \mh = \mh_{\hbox{\footnotesize phot}} - \mh_{\hbox{\footnotesize spec}} \approx$ 0.5 dex.
This result would support the spectroscopic measurements of ZO04 for NGC~6528, which give 
$\feh = $-0.10$\pm$0.2 and a low $\alpha$-enhancement, i. e. \afe $\approx$ 0.1, and the 
findings of \citet{carretta01}, based on high-resolution spectroscopy of RHB stars, 
which give $\feh =$ 0.07$\pm$0.01, with a modest $\alpha$-enhancement, \afe $\approx$ 0.2.

The smoothed photometric metallicity distribution of proper-motion selected and differential 
reddening corrected RGs in the Baade's window surrounding NGC~6528, based on 
scaled-solar $[m], \ y - J$ MIC relation, is clearly bimodal, with a main 
peak at \feh $\approx -0.25$ and a secondary one at \feh $\approx$ 0.5. Similar distributions 
are obtained when adopting the other visual--NIR MIC relations, with average peaks at 
\feh $\approx -0.2$ and \feh $\approx$ 0.55 ($[m],\ y - J$, $[m],\ y - H$ and $m_1,\ y - K$ relations),
and \feh $\approx -0.25$ and \feh $\approx$ 0.4 ($m_1,\ y - J$, $m_1,\ y - H$ , and $m_1,\ y - K$ relations).
By applying the $\alpha$-enhanced MIC relations we obtain a metallicity distribution shifted 
by almost 1 dex towards the metal-rich regime. 

The shape of our photometric metallicity distribution is quite similar to the shape of the spectroscopic 
distribution derived by ZO08 and \citet{hill11} based on a
sample of $\approx$ 200 RGs in the Baade's window. The metal-rich peak of the photometric 
distribution is slightly shifted ($\approx$ 0.15 - 0.2 dex) towards the metal-rich regime. The shift
could be due to the effect of the molecular $CN/CH/NH$-bands on the $m_1$ index, 
so the RGs mimic to be more metal-rich than what they really are. Moreover, our MIC
relations are valid in the metallicity range $-2.3 <$ \feh $<$ 0.3, and can be safely adopted in the 
range $-2.5 <$ \feh $<$ 0.5, taking into account the current uncertainties in spectroscopic abundances 
and in the metallicity scale. The metal-rich peak of our distribution is at the edge of the metallicity range 
covered by our MIC relations. 

It is worth noticing that our photometric metallicity distribution for the RGs in the Baade's window is also 
in fairly good agreement with the spectroscopic MDF obtained by the ARGOS
survey for RGs in the Galactic bulge at latitude $b = -5^{\circ}$ \citep[see panel a) of their Fig.~11]{ness13}, 
with the spectroscopic MDF derived by \citet{bensby13} for a sample of 58 micro-lensed bulge dwarfs, 
and with the MDF obtained by \citet{utten} for a sample of $\sim$ 400 RGs in the a region 
of the bulge centered at $(l,b) = (0^{\circ}, -10^{\circ})$.

The above findings indicate that \stromc--NIR photometry can provide solid constraints
on the metallicity distribution not only of cluster RGs but also of field and halo bulge RGs. 
The key advantage of the current approach is that it can be applied to a large sample 
of field RGs located along different line of sight. The main drawback is that it requires 
accurate and deep photometry in at least three \strom bands ($v,b,y$) and at least one NIR 
band. However, accurate NIR photometry is now available for a significant fraction of the 
Galactic bulge (VISTA, Saito et al.\ 2012). An extensive use of the current approach is 
hampered by the fact that current wide field imager at 4--8m class telescopes are not 
equipped with high-quality \strom filters. It goes without saying that a medium--large 
\strom survey of the low-reddening regions of the Galactic bulge can have a substantial 
impact on our understanding of the bulge stellar populations. This means more quantitative 
constraints on the formation and evolution of the bulge and its chemical enrichment.

\begin{acknowledgements}
We thank Adriano Pietrinferni for his help with ZAHB models, and
Elizabeth Fraser for helping us with the English.
We acknowledge the referee for his/her pertinent comments and suggestions that helped us to improve the 
content and the readability of the manuscript.
Support for this work has been provided by the IAC (grant 310394), and the Education and Science 
Ministry of Spain (grants AYA2007-3E3506, and AYA2010-16717).
This work was partially supported by PRIN INAF 2011 ``Tracing the formation
and evolution of the Galactic halo with VST'' (PI: M. Marconi) and by
PRIN--MIUR (2010LY5N2T) ``Chemical and dynamical evolution of the
Milky Way and Local Group galaxies'' (PI: F. Matteucci).
FM is thankful for the financial support from FONDECYT for project 3140177 and from the Chilean BASAL 
Centro de Excelencia en Astrof\'isica y Tecnolog\'ias Afines (CATA) grant PFB-06/2007.
APM acknowledges the financial support from the Australian Research Council through Discovery Project grant DP120100475.
\end{acknowledgements}


\end{document}